\documentclass[10pt, nolongbibliography,groupedaddress,showpacs,showkeys,amsmath,amssymb,eqsecnum,aps,prd,nofootinbib]{revtex4-2}

\usepackage{lmodern}

\usepackage[]{graphicx}
\usepackage[]{graphics}
\usepackage{amsmath}
\usepackage{tensor,mathtools, amsthm} 
\usepackage{epsf}                                                                                           
\usepackage{color}                   
\usepackage{verbatim}
\usepackage[mathscr]{euscript}

\usepackage{hyperref}

\newcommand{\be}{\begin{equation}}

\newcommand{\ee}{\end{equation}}

\begin{document}


 \author{F. T. Brandt}  
 \email{fbrandt@usp.br}
 \affiliation{Instituto de F\'{\i}sica, Universidade de S\~ao Paulo, S\~ao Paulo, SP 05508-090, Brazil}

\author{J. Frenkel}
\email{jfrenkel@if.usp.br}
\affiliation{Instituto de F\'{\i}sica, Universidade de S\~ao Paulo, S\~ao Paulo, SP 05508-090, Brazil}

\author{S. Martins-Filho}    
\email{sergiomartinsfilho@usp.br}
\affiliation{Instituto de F\'{\i}sica, Universidade de S\~ao Paulo, S\~ao Paulo, SP 05508-090, Brazil}

 \author{D. G. C. McKeon}
 \email{dgmckeo2@uwo.ca}
 \affiliation{
 Department of Applied Mathematics, The University of Western Ontario, London, Ontario N6A 5B7, Canada}
 \affiliation{Department of Mathematics and Computer Science, Algoma University, Sault Ste.~Marie, Ontario P6A 2G4, Canada}

\title{Quantization of Einstein-Cartan Theory in the First Order Form}

\date{\today}

\begin{abstract}
We consider the Einstein-Cartan theory with the tetrad $e^a_\mu$ and spin connection $\omega_{\mu a b}$ taken as being independent fields. Diffeomorphism invariance and local Lorentz invariance result in there being two distinct gauge transformations in this approach, and consequently two ghost fields arise when employing the usual Faddeev-Popov quantization procedure. Our choice of gauge fixing retains the gauge invariances of the background field. We show that the gauge algebra is closed even in the presence of torsion, and the resulting BRST invariance can be found for the effective action. We also derive the Slavnov-Taylor identities, which reflect the BRST symmetries of this theory.
\end{abstract}


\maketitle

\section{Introduction}\label{sec1} 

The Einstein-Cartan theory, first proposed by Cartan in 1922 \cite{Cartan}, is an extension of general relativity which enables 
space-time to have curvature and torsion. In 1928, Einstein introduced torsion in his unified theory of gravity and electromagnetism \cite{Einstein:1948kr, Yepez:2011bw}. In the 1960s, assuming that matter-energy is the source of curvature and spin is the source of torsion, Kibble and Sciama \cite{Kibble:1961ba, Sciama:1964wt} developed a consistent theory of gravitation with curvature and torsion.

In general relativity (GR) the fundamental entity is the metric $g_{\mu\nu}(x)$, but this field cannot be directly coupled to a Dirac field $\psi(x)$. The most convenient way to couple the graviton to spinors utilizes the tetrad field $e^a_\mu(x)$, which is the ``square root'' of the metric \cite{Weyl:1950xa,Utiyama:1956sy}. One of the actions for this field is the Einstein-Cartan (EC) action \cite{Hehl:1976kj, Shapiro:2001rz, Hammond:2002rm, Cai:2015emx,Karananas:2021zkl}, which 
differs from that in GR \cite{Deser:1976ay, Nakanishi:1982ag, Buchbinder:2021wzv} by being formulated in the Riemann-Cartan geometry, that embodies a local Lorentz symmetry. Moreover, the EC theory introduces an affine connection with an antisymmetric part, which can describe a space-time with torsion. This may be able to avoid the GR problem of the singularity of the Big-Bang \cite{Poplawski:2011jz}.
This theory may be regarded as a gauge theory of the Poincar\'e group, consisting of translations and Lorentz transformations, which makes gravity more akin to the other fundamental interactions \cite{Hehl:2013qga, Hehl:2023khc}. 
The first order form of the EC theory involves apart from the tetrad $ e_{\mu}^{a} $ an independent  spin connection $\omega_{\mu a b}$ \cite{Castellani:1981ue, DiStefano:1982va}, which are gauge fields 
associated with the invariances under local translations and local Lorentz transformations.

In the presence of spinning matter, the GR and EC theories differ at the classical level. 
These theories also differ at the quantum level, since in this case there is no relation between the quantum tetrad and the spin connection fields. In this paper, we quantize the EC theory by employing gauge fixings for both of these fields. The gauge invariances present in the background field are unbroken. Both of these gauge transformations result in the introduction of Faddeev-Popov (FP) ghosts. We assert that the full effective action (the sum of the classical EC action, the gauge fixing terms, and the ghost terms) possesses BRST symmetries \cite{Becchi:1974md,Tyutin:1975qk, Lavrov:1997xk} compatible with a closed gauge algebra. 

The paper is organized as follows. In Sec.\ II we reproduce some basic relations obtained in the EC theory. An analysis of the quantization of a general theory with two gauge invariances is given in Sec.\ III. This treatment is applied to the EC theory in first-order form, in Sec.\ IV.  In the framework of the background field method, we show that the algebra of the gauge transformations is closed, which ensures the existence of  BRST invariances consistent with both gauge transformations. The corresponding Slavnov-Taylor identities involving the generating functional of proper Green's functions are examined in Sec. V. A  brief discussion of the results is given in Sec.\ VI. Some relevant details of the calculations are presented in the appendices. 

\section{The Classical EC Action}\label{sec2} 
The EC gravity is characterized by an affine connection $ \tensor{\Gamma}{_{\mu \nu}^{\lambda} }$, whose antisymmetric part defines the torsion tensor
\begin{equation}\label{eq:e4ab}
    \tensor{S}{_{\mu \nu}^{\lambda}}= \frac{1}{2} \left ( \tensor{\Gamma}{_{\mu \nu}^{\lambda}} - \tensor{\Gamma}{_{\nu \mu}^{\lambda}}\right ).
\end{equation}
Imposing the metricity of the covariant derivative: $ g_{\mu \nu; \lambda} =0$, one can uniquely express the affine connection in terms of the Levi-Civita connection $ \{ \begin{smallmatrix} \lambda \\ \mu \nu \end{smallmatrix}\} $ and of the contorsion tensor $ \tensor{K}{_{\mu \nu}^{\lambda}} $ as
\be\label{e4}
\tensor{\Gamma}{_{\mu \nu}^{\lambda}} = 
\begin{Bmatrix}
    \lambda \\
    \mu \nu
\end{Bmatrix}
 - \tensor{K}{_{\mu \nu}^{\lambda}},
\ee
where 
\begin{equation}\label{eq:e4a}
\begin{Bmatrix}
    \lambda \\
    \mu \nu 
\end{Bmatrix}
=
    \frac{1}{2} g^{\lambda\rho}\left(
g_{\rho\mu,\nu}+g_{\rho\nu,\mu}-g_{\mu\nu,\rho}
\right)
\end{equation}
and
\begin{equation}\label{eq:e4b}
    K_{\mu \nu \lambda} = S_{\nu \lambda \mu} + S_{\mu \lambda \nu} - S_{\mu \nu \lambda}.
\end{equation}
It is worth noticing that the torsion tensor $ S_{\mu \nu \lambda} $ is antisymmetric in the first two indices, while the contorsion tensor is antisymmetric in the last two indices.

The EC action involves the tetrad field $e^a_\mu(x)$ that is related to the metric as 
\be\label{e1}
g_{\mu\nu}(x) = \eta_{ab} e^a_\mu(x) e^b_\nu(x);\;\;\;\; (\eta_{ab}=\text{diag}(+---)).
\ee
 If the total covariant derivative of $e^a_\mu(x)$ vanishes,
(comma denotes partial derivatives and semicolons the usual covariant derivative, see Eq.~\eqref{eq:A1}.)
\begin{eqnarray}\label{e2}
    {\cal D}_\lambda e^a_\mu &=&  e^a_{\mu ; \lambda}+ \omega^{\;a\;}_{\lambda\; b} e^b_\mu = 0
\nonumber \\
                             &=&  e^a_{\mu , \lambda} - \tensor{\Gamma}{_{\lambda\mu}^\sigma} e^a_\sigma + \omega^{\;a\;}_{\lambda\; b} e^b_\mu   = 0,
\end{eqnarray}
where $ \omega_{\mu ab} $ is the spin connection,
then this equation implies the metricity condition $ g_{\mu \nu ; \lambda} =0$.
Using Eq.~\eqref{e4} and defining the antisymmetric tensor $ K_{\mu ab} = 1/2(e_{a}^{\nu} e_{b}^{\lambda} - e_{a}^{\lambda} e_{b}^{\nu} ) K_{\mu \nu \lambda}$ one can check that Eq.~\eqref{e2} leads to the following relation:
\be\label{e3}
\omega_{\mu ab} = \frac{1}{2}e^\nu_a\left( e_{b\nu, \mu } -  e_{b\mu, \nu}\right) + 
\frac{1}{2} e^\sigma_a  e^\lambda_b \left( e_{c\sigma , \lambda } -  e_{c\lambda , \sigma }\right) e^c_\mu
- (a\leftrightarrow b)
+ K_{\mu ab},
\ee
where the spin connection $ \omega_{\mu ab} $ is antisymmetric in the last two indices $ \omega_{\mu ab} = - \omega_{\mu ba} $.

Using this relation, the EC action in the second order formalism may be written in the form
\be\label{e5}
S_{\text{EC}}= -\frac{1}{16 \pi G} \int \mathop{d^4x} e R(e, \omega)
\ee
where $G$ is Newton's constant,
\be\label{e6}
e = \det e^a_\mu = \sqrt{-\det g_{\mu\nu}}
\ee
and
\be\label{e7}
R(e , \omega) = e^{a\mu} e^{b\nu} R_{\mu\nu ab}(\omega)
\ee
with
\be\label{e8o}
R_{\mu\nu ab} ( \omega)=  \omega_{\nu ab, \mu } -  \omega_{\mu ab, \nu } + \omega_{\mu ap} \omega_{\nu\;b}^{\; p\;}
-\omega_{\nu ap} \omega_{\mu\;b}^{\; p\;} .
\ee

The first order EC action is identical in form to Eq. \eqref{e5}, but now $e_\mu^a$ and $\omega_{\mu a b}$ are treated as independent fields. (This is analogous to the ``Palatini'' form of the Einstein-Hilbert (EH) action in which $g_{\mu\nu}$ and $\tensor{\Gamma}{_{\sigma \beta }^{\lambda} }$ are treated as being independent.) One can show that the solution to the equation of motion for $\omega_{\mu ab}$ that follows from Eqs. \eqref{e5} and \eqref{e8o}, when $ e  \neq 0$, results in Eq. \eqref{e3} \cite{Kiriushcheva:2009tg}. 
This establishes that at the classical level, the first and second order formulations of EC theory are equivalent.

The first order action possesses, first of all, diffeomorphism invariance, so that if ${x}^{\prime\mu} = x^\mu + \epsilon^\mu(x)$, then for a vector field $V_\mu(x)$, we find for infinitesimal functions $\epsilon^\mu(x)$,

\begin{subequations}\label{e8}
\be\label{e8a}
\delta_\epsilon V_\mu(x) = 
-\epsilon^\lambda V_{\mu , \lambda} - V_\lambda \tensor{ \epsilon}{^\lambda_{,\mu}}
\ee
\be\label{e8b}
\delta_\epsilon V^\mu(x) =
 - \epsilon^{\lambda} \tensor{V}{^{\mu}_{, \lambda}} + V^{\lambda} \tensor{\epsilon}{^{\mu}_{, \lambda} }.
\ee
\end{subequations}
Local Lorentz invariance in which
\begin{subequations}\label{e9}
\be\label{e9a}
         {e^\prime}^a_\mu(x) = \Lambda^a_{\;b}(x) e^b_\mu(x);\;\;\; (\Lambda^a_{\; b} \eta_{ac} \Lambda^c_{\; d} = \eta_{bd})
\ee
results in         
\be\label{e9b}
\delta_\lambda e^a_\mu 
= \lambda^a_{\;b} e^b_\mu;\;\;\; (\lambda_{ab} = -\lambda_{ba})
\ee
\end{subequations}
if $\Lambda^a_{\; b} = \delta^a_{b} + \lambda^a_{\; b}$ when $\lambda$ is infinitesimal. 

We now consider the Dirac matrices $\gamma^a$ with $\{\gamma^a,\gamma^b \}=2\eta^{ab} I$ and then define
\be\label{e10a}
\sigma^{ab} = \frac{1}{4} [\gamma^a,\gamma^b].
\ee
If a spinor field transforms as
\be\label{e12}
\psi^\prime = U(\Lambda) \psi;\;\;\; (U\approx 1+\frac{1}{2} \sigma^{ab} \lambda_{ab})
\ee
then
\be\label{e13}
D_\mu(\omega^\prime)\psi^\prime = UD_\mu(\omega)\psi;\;\;\; (D_\mu\equiv\partial_\mu+\frac{1}{2} \sigma^{ab}\omega_{\mu ab})
\ee
provided
\be\label{e14}
\delta_\lambda \omega_{\mu ab} = -\lambda_{ab, \mu } + \omega_{\mu\; b}^{\; p\;}
\lambda_{ap} + \omega_{\mu a}^{\; \;\;p\;} \lambda_{bp} .
\ee
Eqs.~\eqref{e8}-\eqref{e14} establish the two gauge invariances of the Dirac action in the presence of gravity:
\begin{equation}\label{eq:e14a}
    S_{\text{D}} = \int \mathop{d^{4}x} e \left [ \frac{i}{2} \left ( \bar{\psi} e_{a}^{\mu} \gamma^{a} D_{\mu} \psi - \overline{D_{\mu} \psi} e^{\mu}_{a} \gamma^{a} \psi\right ) - m \bar{\psi} \psi\right ] . 
\end{equation}
This action yields a spin density which can generate torsion \cite{Hehl:1976kj, Shapiro:2001rz, Hammond:2002rm, Karananas:2021zkl, Cai:2015emx}.

The path integral can now be used to quantize the EC action, with the gauge invariance of Eqs. \eqref{e8}, \eqref{e9} and \eqref{e14} being taken into account through a slightly modified FP procedure  \cite{Faddeev:1967fc,Feynman:1963ax,DeWitt:1967ub,Mandelstam:1962mi}.

\section{Quantization with two gauge invariances}
Consider the general case of two fields $\phi_i$ and $\Phi_I$ that undergo the infinitesimal gauge transformations
\begin{subequations}\label{eq:15}
    \begin{align}\label{eq:15a}
        \delta \phi_{i} ={}& r_{ij} ( \phi ) \theta_{j} + r_{iJ} ( \phi ) \Theta_{J},
      \\  \label{eq:15b}
      \delta \Phi_{I} ={}& R_{Ij} ( \Phi ) \theta_{j} + R_{IJ} ( \Phi ) \Theta_{J} .
    \end{align}
\end{subequations}
To these fields, we apply the gauge conditions 
\begin{equation}\label{eq:16}
    f_{ij} \phi_{j} = 0 = F_{IJ} \Phi_{J} .
\end{equation}
If the classical action $ S_{\text{cl} } = \int \mathop{d x} \mathcal{L}_{ \text{cl} } ( \phi_{i} , \Phi_{I} )$ is invariant under the transformation of Eq.~\eqref{eq:15}, then the path integral 
\begin{equation}\label{eq:17}
    \int \mathop{\mathcal{D} \phi_{i}} \mathop{\mathcal{D} \Phi_{I}} \exp{i S_{ \text{cl} }}
\end{equation}
is modified by introduction of a constant factor 
\begin{equation}\label{eq:18}
    \int \mathop{\mathcal{D} \theta_{i}} \mathop{\mathcal{D} \Theta_{I}} \delta \left [ \begin{pmatrix}
            f_{ij} & 0
            \\ 0 & F_{IJ}
\end{pmatrix}
\left ( \begin{pmatrix}
        \phi_{j} \\
        \Phi_{J}
\end{pmatrix}
+ \begin{pmatrix}
    r_{jk} & r_{jK} 
    \\ R_{Jk} & R_{JK}
\end{pmatrix}
\begin{pmatrix}
    \theta_{k} \\
    \Theta_{K}
\end{pmatrix}
\right ) - \begin{pmatrix}
    p_{i} \\
    P_{I}
\end{pmatrix}
\right ] 
\det \begin{pmatrix}
    f_{ij} r_{jk} & f_{ij} r_{jK} \\
    F_{IJ} R_{Jk} & F_{IJ} R_{JK}
\end{pmatrix}.
\end{equation}
The determinant in Eq.~\eqref{eq:18} can be exponentiated using Fermionic ghost fields $ c_{i} , \bar{c}_{i} , C_{I}, \bar{C}_{I} $ and we find 
\begin{equation}\label{eq:19}
\det \begin{pmatrix}
    f_{ij} r_{jk} & f_{ij} r_{jK} \\
    F_{IJ} R_{Jk} & F_{IJ} R_{JK}
    \end{pmatrix} = \int \mathop{\mathcal{D} c_{i}} \mathop{\mathcal{D} \bar{c}_{i}} \mathop{\mathcal{D} C_{I}} \mathop{\mathcal{D} \bar{C}_{I}} \exp{i \int \mathop{d^{}x} \begin{pmatrix}
    \bar{c}_{i} & \bar{C}_{I}
\end{pmatrix}
\begin{pmatrix}
    f_{ij} r_{jk} & f_{ij} r_{jK} \\
    F_{IJ} R_{Jk} & F_{IJ} R_{JK}
\end{pmatrix}
\begin{pmatrix}
    c_{k} \\
    C_{K}
\end{pmatrix}
} .
\end{equation}
The gauge fixing of Eq.~\eqref{eq:16} can also be incorporated into the effective action by inserting into Eq.~\eqref{eq:17} the constant factor 
\begin{equation}\label{eq:20}
    \int \mathop{\mathcal{D} p_{i}} \mathop{\mathcal{D} P_{I}} \int \mathop{\mathcal{D} b_{i}} \mathop{\mathcal{D} B_{I}} \exp{i \int \mathop{d x} \left ( \alpha b_{i} b_{i} + \beta B_{I} B_{I} - b_{i} p_{i} - B_{I} P_{I}\right )},
\end{equation}
where $b_{i} $ and $ B_{I} $ are Nakanishi-Lautrup fields \cite{Nakanishi:1966zz, Lautrup:1967zz} which can be used to build an off-shell nilpotent BRST transformation. 
Once Eqs.~\eqref{eq:18} and \eqref{eq:20} are inserted into Eq.~\eqref{eq:17}, a gauge transformation of the form of Eq.~\eqref{eq:15} with $ \theta_{i} , \Theta_{I} $ replaced by $- \theta_{i} , - \Theta_{I} $ is made. The $ \delta$-function appearing in Eq.~\eqref{eq:18} can then be used to integrate over $ p_{i} $ and $ P_{I} $. We are then left with the effective Lagrangian
\begin{equation}\label{eq:21}
    \begin{split}
        \mathcal{L}_{\text{eff}} ={}& \mathcal{L}_{ \text{cl}} + \alpha b_{i} b_{i} - b_{i} f_{ij} \phi_{j} + \beta B_{I} B_{I} - B_{I} F_{IJ} \Phi_{J} + \bar{c}_{i} f_{ij} r_{jk} ( \phi ) c_{k} \\ & +  \bar{C_{I}} F_{IJ} R_{JK} ( \Phi ) C_{K} + \bar{c}_{i} f_{ij} r_{jK} ( \phi ) C_{K} + \bar{C}_{I} F_{IJ} R_{Jk} ( \Phi ) c_{k} .
\end{split}
\end{equation}
The integral over $ \theta_{i} $ and $ \Theta_{I} $ in Eq.~\eqref{eq:18} now become an overall multiplicative factor in the path integral, and what remains in the path integral is well defined. If we had tried to eliminate the gauge degrees of freedom associated with the two gauge transformations of Eq.~\eqref{eq:15} by applying the FP procedure to first eliminate the transformation of $ \theta_{i} $ and then using a second application of this procedure to subsequently eliminate $ \Theta_{I} $, then $ \mathcal{L}_{\text{eff}} $ would in this case retain dependency on $ \Theta_{I} $ that would reside in the gauge fixing term.

As a consequence of having eliminated $ \theta_{i} $ and $ \Theta_{I} $ together, in Eq.~\eqref{eq:21} there are interaction terms involving $ (\bar{C}_{I}, \, c_{k})$ and $ (\bar{c}_{i} , \, C_{K})$ so that there are vertices involving both ghosts.

The question of having BRST invariance \cite{Becchi:1974md, Tyutin:1975qk} or its generalization \cite{deWit:1978hyh,VanNieuwenhuizen:1981ae,Batalin:1983ggl} can now be examined. The first step is to consider, as usual from Eq.~\eqref{eq:15}, the transformation
\begin{subequations}\label{eq:22}
    \begin{align}
        \label{eq:22a}
        \delta \phi_{i} ={}& \left(r_{ij} ( \phi ) c_{j} + r_{iJ} ( \phi ) C_{J}\right) \eta, 
       \\ \label{eq:22b}
       \delta \Phi_{I} ={}& \left(R_{Ij} ( \Phi ) c_{j} + R_{IJ} ( \Phi ) C_{J}\right) \eta 
    \end{align}
\end{subequations}
where $ \eta $ is a constant Grassmann scalar. We then take, from Eq.~\eqref{eq:21}, \begin{align} \label{eq:23}
    \delta b_{i} ={}& \delta B_{I} = 0, \\ \label{eq:24}
    \delta \bar{c}_{i} ={}& - b_{i} \eta \quad \text{and} \quad \delta \bar{C}_{I} = - B_{I} \eta.
\end{align}
We now require that \begin{subequations}\label{eq:25}
    \begin{align}
        \label{eq:25a}
        \mathop{\delta} [r_{jk} ( \phi ) c_{k} + r_{jK} ( \phi ) C_{K}] ={}&0 
       \\ \nonumber 
       \shortintertext{and}
       \label{eq:25b} 
       \mathop{\delta} [R_{Jk} ( \Phi ) c_{k} +R_{JK} ( \Phi ) C_{K}] ={}& 0.
    \end{align}
\end{subequations}

Eqs.~\eqref{eq:22} and \eqref{eq:25} together lead to the two conditions
\begin{subequations} \label{eq:26}
\begin{gather}\label{eq:26a}
    \left ( r_{jk} \delta c_{k} + r_{jK} \delta C_{K} \right ) + \left [ \left ( \frac{\partial r_{jk}}{\partial \phi_{l}} c_{k} + \frac{\partial r_{jK}}{\partial \phi_{l}} C_{K}\right ) (r_{lm} c_{m} + r_{lM} C_{M} ) \eta\right ] =0, \\ \label{eq:26b}
    \left ( R_{Jk} \delta c_{k} + R_{JK} \delta C_{K} \right ) + \left [ \left ( \frac{\partial R_{Jk}}{\partial \Phi_{L}} c_{k} + \frac{\partial R_{JK}}{\partial \Phi_{L}} C_{K}\right ) (R_{Lm} c_{m} + R_{LM} C_{M} ) \eta\right ] =0.
\end{gather}
\end{subequations}
As $ c_{i} $ and $ C_{I} $ are Grassmann, we see that if the gauge transformations of Eq.~\eqref{eq:15} satisfy a ``closed'' algebra so that \begin{subequations}\label{eq:27}
    \begin{align} \label{eq:27a}
        \frac{1}{2} \left ( \frac{\partial r_{jk}}{\partial \phi_{l}} r_{lm} - \frac{\partial r_{jm}}{\partial \phi_{l}} r_{lk}\right ) ={}& r_{jl} A_{km}^{l} + r_{jL} B_{km}^{L},\\ \label{eq:27b}
        \frac{1}{2}\left ( \frac{\partial r_{jK}}{\partial \phi_{l}} r_{lM} - \frac{\partial r_{jM}}{\partial \phi_{l}} r_{lK}\right ) ={}& r_{jl} A_{KM}^{l} + r_{jL} B_{KM}^{L}, \\ \label{eq:27c}
        \frac{\partial r_{jk}}{\partial \phi_{l}} r_{lM} - \frac{\partial r_{jM}}{\partial \phi_{l}} r_{lk} ={}& r_{jl} A_{kM}^{l} + r_{jL} B_{kM}^{L} 
    \end{align}
\end{subequations}
Then, Eq.~\eqref{eq:26a} can be satisfied provided \begin{subequations}\label{eq:28}
    \begin{align} \label{eq:28a}
        \delta c^{l} ={}&- \left (  A_{km}^{l} c_{k} c_{m} +  A_{KM}^{l} C_{K} C_{M} + A_{kM}^{l} c_{k} C_{M}\right ) \eta, \\ \label{eq:28b}
        \delta C^{L} ={}&- \left (  B_{km}^{L} c_{k} c_{m} +  B_{KM}^{L} C_{K} C_{M} + B_{kM}^{L} c_{k} C_{M}\right ) \eta
    \end{align}
\end{subequations}
Expressions for $ \delta c^{\mu} $ and $ \delta C^{ab} $ analogous to those in Eq.~\eqref{eq:28}  can be obtained from Eq.~\eqref{eq:26b}. For consistency, $ \delta c^{\mu} $ and $ \delta C^{ab} $ should be unique, and so the matrices $A$ and $B$ derived from Eq.~\eqref{eq:26a} and Eq.~\eqref{eq:26b} must be the same. 
This requires certain functional relations between the matrices $r$ and $R$ in Eq.~\eqref{eq:15} (see Eqs.~\eqref{eq:c3} and \eqref{eq:c4}).


In Refs. \cite{deWit:1978hyh,VanNieuwenhuizen:1981ae,Batalin:1983ggl, Tyutin:1986ts} the presence of a BRST transformation when the gauge transformations form an ``open'' algebra is discussed, where an open algebra is that of Eq.~\eqref{eq:27} that is satisfied only on the ``mass shell'' (i.e., when the classical field equations are satisfied). 
We can also construct an off-shell nilpotent BRST transformation for open algebras by introducing auxiliary fields \cite{Henneaux:1988ej}, which is analogous to the introduction of the Nakanishi-Lautrup fields.

Next, we will consider the BRST quantization of the  first-order EC action given in Eq.~\eqref{e5}. To this end, it is convenient to use the background field method which requires an invertible background metric with $ \det \bar{g}_{\mu \nu} \neq 0$ \cite{Deser:2006db}. This allows for a comparison of this approach with that used in the second order formulation of the EC theory. We note here that the first order formalism contains another phase with a non-invertible metric, which was first studied in Ref. \cite{Tseytlin:1981ks}.

\section{Quantization of the First-Order EC action}\label{section:QFOEC}

We now identify the fields $ \phi_{i} $ and $ \Phi_{I} $ of Eq.~\eqref{eq:15} with $ e^a_\mu $ and $ \tensor{\omega}{_{\mu ab}}$ respectively. By Eqs.~\eqref{e8}, \eqref{e9} and \eqref{e14}
these undergo the gauge transformations \begin{subequations}\label{eq:30}
    \begin{align} \label{eq:30a}
        \delta e^a_\mu ={}&- \epsilon^{\lambda}  e^a_{\mu, \lambda } - e^{a}_{\lambda}  \tensor{\epsilon}{^{\lambda}_{, \mu} } + \tensor{\lambda}{^{a}_{b}} e^{b}_{\mu}   \\ \shortintertext{and}
        \label{eq:30b} \delta \omega_{\mu ab} ={}& - \epsilon^{\lambda}  \omega_{\mu ab , \lambda } - \omega_{\lambda ab} \tensor{\epsilon}{^{\lambda}_{, \mu} }-  \lambda_{ab , \mu } + \tensor{\lambda}{_{a}^{p}} \omega_{\mu p b} + \tensor{\lambda}{_{b}^{p}} \omega_{\mu ap} . 
    \end{align}
\end{subequations}
We now introduce a background field $ \bar{e}_{\mu}^{a} $ for $ e_{\mu}^{a} $ and $ \bar{\omega}_{\mu ab} $ for $\omega_{\mu ab }$ so that 
\begin{subequations} \label{eq:31}
\begin{align}\label{eq:31a}
    e_{\mu}^{a} = \bar{e}_{\mu}^{a} + q_{\mu}^{a} \\ \shortintertext{and}
   \label{eq:31b}
   \omega_{\mu ab} = \bar{\omega}_{\mu ab} + Q_{\mu ab}
\end{align}
\end{subequations}
with $ q_{\mu}^{a} $ and $ Q_{\mu ab}$ being quantum \cite{DeWitt:1967ub,tHooft:1975uxh}.
 Following Refs. \cite{Abbott:1980hw, Abbott:1981ke}, we will choose gauge fixing terms that respect the symmetries of Eq.~\eqref{eq:30} for the background fields $ \bar{e}_{\mu}^{a} $ and $\bar{\omega}_{\mu ab}$, but break the symmetries \begin{subequations}\label{eq:32}
    \begin{align} \label{eq:32a}
        \delta \bar{e}_{\mu}^{a} ={}&0, \\ \label{eq:32b}
        \delta q_{\mu}^{a} ={}& - \epsilon^{\lambda} e_{\mu , \lambda }^{a} - e_{\lambda}^{a}  \tensor{\epsilon}{^{\lambda}_{, \mu} }+ \tensor{\lambda}{^{a}_{b}}e_{\mu}^{b}
        \shortintertext{and} \label{eq:32c}
        \delta \bar{\omega}_{\mu ab} ={}&0,\\
        \label{eq:32d} 
        \delta Q_{\mu ab} ={}& 
        - \epsilon^{\lambda} \omega_{\mu ab , \lambda } - \omega_{\lambda ab}  \tensor{\epsilon}{^{\lambda}_{, \mu} }-  \lambda_{ab , \mu } + \tensor{\lambda}{_{a}^{p}} \omega_{\mu p b} + \tensor{\lambda}{_{b}^{p}} \omega_{\mu ap} . 
    \
    \end{align}
\end{subequations}
If we define a covariant derivative using the background metric $ \bar{g}_{\mu \nu} $ where 
\begin{equation}\label{eq:33}
    \bar{g}_{\mu \nu} = \eta_{a b} \bar{e}_{\mu}^{a} \bar{e}_{\nu}^{b}
\end{equation}
to be 
\begin{equation}\label{eq:34}
    V_{\mu ; \bar{\lambda}} = V_{\mu , \lambda} - \tensor{\bar{\Gamma}}{_{\lambda \mu }^{\sigma}} V_{\sigma} 
\end{equation}
with $ \tensor{\bar{\Gamma}}{_{\mu \lambda}^{\sigma}} $ defined in Eq.~\eqref{e4} with $ \bar{g}_{\mu \nu} $ replacing $ g_{\mu \nu} $, then Eqs.~\eqref{eq:32b} and \eqref{eq:32d} can be written in a covariant form \begin{subequations}\label{eq:35}
    \begin{align} \label{eq:35a}
        \delta q_{\mu}^{a} ={}&- \epsilon^{\lambda} e_{\mu; \bar{\lambda}}^{a} - e_{\lambda}^{a} \tensor{\epsilon}{^{\lambda}_{; \bar{\mu}}} +2 \tensor{\bar{S}}{_{\mu \lambda}^{\beta}} e_{\beta}^{a} \epsilon^{\lambda} + \tensor{\lambda}{^{a}_{b}}e_{\mu}^{b},\\
        \label{eq:35b}
        \delta Q_{\mu ab} ={}& - \epsilon^{\lambda} \omega_{\mu ab; \bar{\lambda}} - \omega_{\lambda ab} \tensor{\epsilon}{^{\lambda}_{; \bar{\mu}}} +2 \tensor{\bar{S}}{_{\mu \lambda}^\beta} \omega_{\beta ab} \epsilon^{\lambda} -  \lambda_{ab , \mu } + \tensor{\lambda}{_{a}^{p}} \omega_{\mu pb} + \tensor{\lambda}{_{b}^{p}} \omega_{\mu ap},
    \end{align}
\end{subequations}
where $ \tensor{\bar{S}}{_{\mu \nu}^{\lambda}}$ is defined in Eq.~\eqref{eq:e4ab} with $ \tensor{\bar{\Gamma}}{_{\mu \nu}^{\lambda}} $ replacing $ \tensor{\Gamma}{_{\mu \nu}^{\lambda}}$.

To fix the gauge transformations of Eqs.~\eqref{eq:32a}, \eqref{eq:32c} and \eqref{eq:35}, we choose the simple gauge fixing conditions \begin{subequations}\label{eq:36}
    \begin{align} \label{eq:36a}
        \eta_{ab} \left ( \bar{e}_{\mu}^{a} q_{\nu}^{b} + q_{\mu}^{a} \bar{e}_{\nu}^{b}\right )^{; \bar{\mu}} 
        ={}&0 
        \\ \label{eq:36b}
        \tensor{Q}{_{\mu ab}^{; \bar{\mu}}} ={}&0.
    \end{align}
\end{subequations}
With these choices, we do not break diffeomorphism invariance in the background field provided indices are raised and lowered using $ \bar{g}_{\mu \nu} $.

Following the  development that led to Eq.~\eqref{eq:21}, we find that the effective Lagrangian that follows from the first-order EC action in Eq.~\eqref{e5} and from the Dirac action Eq.~\eqref{eq:e14a} is 
\begin{equation}\label{eq:37}
    \begin{split}
    \mathcal{L}_{\text{eff}} ={}&
    \mathcal{L}_{ \text{EC} } + \mathcal{L}_{\text{D}} + \bar{e}\Bigl\{\alpha b^{\nu} b_{\nu} - b^{\nu} \left[ \eta_{ab} \left ( \bar{e}_{\mu}^{a} q_{\nu}^{b} + q_{\mu}^{a} \bar{e}_{\nu}^{b}\right )^{; \bar{\mu}}\right] 
    + \beta B^{ab} B_{ab} -B^{ab} ( \tensor{Q}{_{\mu ab}^{; \bar{\mu}}}) \\ 
                                & + \bar{c}^{\nu}    \eta_{ab}  \left[ \bar{e}_{\mu}^{a} \left ( - c^{\lambda} e_{\nu ; \bar{\lambda}}^{b} - e_{\lambda}^{b} c_{\  ; \bar{\nu}}^{\lambda} + 2 \tensor{\bar{S}}{_{\nu \beta}^{\lambda}} e_{\lambda}^{b} c^{\beta}  \right ) + \bar{e}_{\nu}^{b} \left ( - c^{\lambda} e_{\mu ; \bar{\lambda}}^{a} - e_{\lambda}^{a} c_{\  ; \bar{\mu}}^{\lambda} + 2 \tensor{\bar{S}}{_{\mu \beta}^{\lambda}} e_{\lambda}^{a} c^{\beta} \right )\right]^{; \bar{\mu}} \\
                                & \quad + \bar{C}^{ab} \left [ -  C_{ab , \mu } + \tensor{C}{_{a}^{p}} \omega_{\mu pb} + \tensor{C}{_{b}^{p}} \omega_{\mu ap}\right ]^{; \bar{\mu} } + \bar{c}^{\nu} \left \{ \eta_{ab} \left [ \bar{e}_{\mu}^{a} \left ( \tensor{C}{^{b}_{p}}e_{\nu}^{p} \right )+ \bar{e}_{\nu}^{b} \left ( \tensor{C}{^{a}_{p}}e_{\mu}^{p}\right )\right ]\right\}^{; \bar{\mu}} \\ 
                                & \quad + 
                            \bar{C}^{ab} \left ( - c^{\lambda} \omega_{\mu ab ; \bar{\lambda}} - \omega_{\lambda ab} c^{\lambda}_{\  ; \bar{\mu}} + 2\tensor{\bar{S}}{_{\mu \lambda}^{\nu}} \omega_{\nu ab} c^{\lambda} \right )^{; \bar{\mu}} \Bigr\},
    \end{split}
\end{equation}
where the factor of $\bar{e}$ is needed to retain the background diffeomorphism invariance of the effective action.
We note that the number of pairs of ghost fields ($ (c_{\mu} , \, \bar{c}_{\mu} )$, $(C_{ab} ,\, \bar{C}_{ab} )$) is the same as the number of gauge functions ($ \epsilon_{\mu} , \lambda_{ab} $).

We note that the gauge \eqref{eq:36b} breaks the local Lorentz invariance, but preserves the diffeomorphism invariance. On the other hand, the gauge \eqref{eq:36a} breaks both the diffeomorphism and local Lorentz invariances, which
leads to a coupling between the ghost fields $ \bar{c}^{\mu} $ and $ C^{ab} $. 
In order to simplify the ghost sector, it is useful to consider instead of \eqref{eq:36a}, the de Donder gauge, which is commonly used in the second-order form \cite{Deser:1974cy,VanNieuwenhuizen:1981ae,Harrison:2013zz}: 
\begin{equation} \label{eq:61old}
\tensor{h}{_{\mu \nu}^{;\bar{\mu}}} - \frac{1}{2}  h_{;\bar{\nu}}=0
\end{equation}
where $ h = h^{\mu \nu}\bar{g}_{\mu\nu}$ and $ g_{\mu \nu} = \bar{g}_{\mu\nu} + h_{\mu\nu}$. In the first order form, using the Eqs.~\eqref{e1}, \eqref{eq:31a} and \eqref{eq:33}, one can relate the field $h_{\mu\nu}$ to the tetrad fields as 
\begin{equation}\label{eq:62}
    h_{\mu \nu} = \eta_{ab} \left ( \bar{e}_{\mu}^{a} q_{\nu}^{b} + q_{\mu}^{a} \bar{e}_{\nu}^{b} + q_{\mu}^{a} q_{\nu}^{b}\right ) . 
\end{equation}
Although this gauge which breaks the diffeomorphism invariance is more involved than the gauge \eqref{eq:36a}, it preserves the local Lorentz invariance and leads to a diagonalization of the ghost propagator matrix.

We now consider the commutator of two gauge transformations. First examining the portion of Eq.~\eqref{eq:35a} given by 
\begin{equation}\label{eq:38}
    \delta_{\xi} q_{\mu}^{a} = - \xi^{\lambda} e_{\mu ; \bar{\lambda }}^{a} - e_{\lambda}^{a} \tensor{\xi}{^{\lambda}_{; \bar{\mu}}} + 2 \tensor{\bar{S}}{_{\mu \lambda}^{\beta}} e_{\beta}^{a} \xi^{\lambda}  
\end{equation}
we find that (the derivation is shown in Appendix A) 
\begin{equation}\label{eq:39}
    \begin{split}
            \left ( \delta_{\xi} \delta_{\zeta} - \delta_{\zeta} \delta_{\xi}\right ) q_{\mu}^{a} ={}& 
            - (\zeta^{\sigma} \tensor{\xi}{^{\lambda}_{;\bar{\sigma}}}-\xi^{\sigma} \tensor{\zeta}{^{\lambda}_{;\bar{\sigma}}} + 2 \tensor{\bar{S}}{_{\sigma \nu }^{\lambda}} \xi^{\sigma} \zeta^{\nu} )  e_{\mu ; \bar{\lambda}}^{a} -e_{\lambda}^{a} (
\zeta^{\sigma} \tensor{\xi}{^{\lambda}_{;\bar{\sigma}}}-\xi^{\sigma} \tensor{\zeta}{^{\lambda}_{;\bar{\sigma}}}
)_{; \mu}  
        +          2 \tensor{\bar{S}}{_{\mu \lambda}^{\beta}} e_{\beta}^{a} (
\zeta^{\sigma} \tensor{\xi}{^{\lambda}_{;\bar{\sigma}}}-\xi^{\sigma} \tensor{\zeta}{^{\lambda}_{;\bar{\sigma}}})
                                                                                                        \\ & \quad +  
            \xi^{\sigma} \zeta^{\lambda} e_{\alpha}^{a} ( -2\tensor{\bar{S}}{_{\sigma \lambda}^{\alpha}_{; \bar{\mu}}} + 4 \tensor{\bar{S}}{_{\sigma \lambda}^{\beta} } \tensor{\bar{S}}{_{\mu \beta}^{\alpha}})
- 2 \tensor{\bar{S}}{_{\lambda \nu}^{ \beta}} e_{\beta}^{a}   
( \zeta^{\nu} \tensor{\xi}{^{\lambda}_{; \bar{\mu}}}- \xi^{\nu} \tensor{\zeta}{^{\lambda}_{; \bar{\mu}}})
\end{split}
\end{equation}
Let us define
\begin{equation}\label{eq:defOmega}
\Omega^{\lambda} = \zeta^{\sigma} \tensor{\xi}{^{\lambda}_{;\bar{\sigma}}} -\xi^{\sigma} \tensor{\zeta}{^{\lambda}_{;\bar{\sigma}}}  + 2 \tensor{\bar{S}}{_{\sigma \nu}^{\lambda} } \xi^{\sigma} \zeta^{\nu}.
\end{equation}
Then we find that Eq.~\eqref{eq:39} becomes\footnote{For more details, see Appendix A.} 
\begin{equation}\label{eq:44}
    \left ( \delta_{\xi} \delta_{\zeta} - \delta_{\zeta} \delta_{\xi}\right ) q_{\mu}^{a} = - \Omega^{\lambda} e_{\mu ; \bar{\lambda}}^{a} - e_{\lambda}^{a} \tensor{\Omega}{^{\lambda}_{; \bar{\mu}}} 
    +2 \tensor{\bar{S}}{_{\mu \lambda}^{\beta}} e_{\beta}^{a}  \Omega^{\lambda}.
\end{equation}
Thus the commutator of two gauge transformations of the form of Eq.~\eqref{eq:38} is itself a gauge transformation of this form. As outlined in Appendix A, this closure property holds as well in the presence of a general tensor.

Similarly, if 
\begin{equation}\label{eq:45}
    \delta_{\lambda} q_{\mu}^{a} = \tensor{\lambda}{^{a}_{b}}e_{\mu}^{b} 
\end{equation}
(as in the last term of Eq.~\eqref{eq:35a}), then 
\begin{equation}\label{eq:46}
    \begin{split}
        \left ( \delta_{\lambda} \delta_{\rho} - \delta_{\rho} \delta_{\lambda}\right ) q_{\mu}^{a} ={}& \left ( \tensor{\rho}{^{a}_{b}} \tensor{\lambda}{^{b}_{c}}- \tensor{\lambda}{^{a}_{b}} \tensor{\rho}{^{b}_{c}} \right ) e_{\mu}^{c}
        \\ \equiv{}& \tensor{L}{^{a}_{b}}e_{\mu}^{b}
    \end{split}
\end{equation}
so that the commutator of two gauge transformations of the form of Eq.~\eqref{eq:45} is itself a gauge transformation of this form.

We now consider the commutator of the gauge transformations of Eqs.~\eqref{eq:38} and \eqref{eq:45}. To compute this, we note that
\begin{equation}\label{eq:47}
    \delta_{\xi} \lambda_{ab} = - \xi^{\sigma} \lambda_{ab , \sigma} 
\end{equation}
since $ \lambda_{ab} $ is itself a scalar (i.e., only the first term on the right side of Eq.~\eqref{e8} contributes). Similarly, using the fact that $ \delta_{\lambda} $ commutes with Lorentz invariant quantities, we find  
\begin{equation}\label{eq:48}
        \delta_{\lambda} e_{\mu , {\sigma}}^{a} = 
        ( \delta_{\lambda} e_{\mu}^{a} )_{, \sigma} 
        =\tensor{\lambda}{^{a}_{b}}e_{\mu , \sigma}^{b} + \tensor{\lambda}{^{a}_{b , \sigma}} e_{\mu}^{b} .
\end{equation}
It then follows that 
\begin{equation}\label{eq:49}
\left ( \delta_{\lambda} \delta_{\xi} - \delta_{\xi} \delta_{\lambda}\right ) q_{\mu}^{a} = 0,
\end{equation}
which shows that the gauge transformations $ \delta_{\xi} $ and $ \delta_{\lambda} $, 
namely, the diffeomorphism and the local Lorentz invariance, commute.

We now consider the gauge transformations that follows from Eq.~\eqref{eq:35b} \begin{subequations}\label{eq:50}
    \begin{align} \label{eq:50a}
        \delta_{\xi} Q_{\mu ab} ={}& - \xi^{\sigma} \omega_{\mu ab; \bar{\sigma}} - \omega_{\sigma ab} \tensor{\xi}{^{\sigma}_{; \bar{\mu}}} + 2 \tensor{\bar{S}}{_{\mu \nu}^{\lambda}} \omega_{\lambda ab} \xi^{\nu} \\ \shortintertext{and}\label{eq:50b}
        \delta_{\lambda} Q_{\mu ab} ={}&  -  \lambda_{ab , \mu } + \tensor{\lambda}{_{a}^{p}}\omega_{\mu pb} + \tensor{\lambda}{_{b}^{p}} \omega_{\mu ap} .
    \end{align}
\end{subequations}
We can establish that \begin{subequations}\label{eq:51}
    \begin{align} \label{eq:51a}
        \left ( \delta_{\xi} \delta_{\zeta} - \delta_{\zeta} \delta_{\xi}\right ) Q_{\mu ab} ={}&- \Omega^{\lambda} \omega_{\mu ab ; \bar{\lambda}} - \omega_{\lambda ab} \tensor{\Omega}{^{\lambda} _{; \bar{\mu}}} + 2 \tensor{\bar{S}}{_{\mu \nu}^{\lambda}} \omega_{\lambda ab} \Omega^{\nu} \intertext{and} \label{eq:51b}
        \left ( \delta_{\lambda} \delta_{\rho} - \delta_{\rho} \delta_{\lambda}\right ) Q_{\mu ab} ={}& -  L_{ab , \mu } + \tensor{L}{_{a}^{p}} \omega_{\mu pb} + \tensor{L}{_{b}^{p}} \omega_{\mu ap}
    \end{align}
\end{subequations}
so that the commutator of two gauge transformations of the form of Eq.~\eqref{eq:50a} (or Eq.~\eqref{eq:50b}) yields a gauge transformation of the same form.

We also find that the commutator of the gauge transformations $ \delta_{\lambda}$ and $\delta_{\xi} $ vanishes:
\begin{equation}\label{eq:52}
    \left ( \delta_{\lambda} \delta_{\xi} - \delta_{\xi } \delta_{\lambda}\right ) Q_{\mu ab} = 0
\end{equation}
where we used
\begin{equation}\label{eq:53}
        \delta_{\lambda} Q_{\mu ab, {\sigma}} =(\delta_{\lambda} Q_{\mu ab})_{, \sigma} = (-\lambda_{ab, \mu } + \omega_{\mu\; b}^{\; p\;}
\lambda_{ap} + \omega_{\mu a}^{\; \;\;p\;} \lambda_{bp})_{, \sigma}  
\end{equation}
and 
\begin{equation}\label{eq:54}
        \delta_{\xi}  \lambda_{ab, \mu } = 
        \left ( \delta_{\xi} \lambda_{ab}\right )_{, \mu} 
        =
        - \tensor{\xi}{^{\sigma}_{, \mu}} \lambda_{ab, \sigma}  
        - \xi^{\sigma} \lambda_{ab, \sigma \mu }.
\end{equation}
We note that Eqs.~\eqref{eq:48} and \eqref{eq:53} ensure that Eqs.~\eqref{eq:45} and \eqref{eq:50b} are consistent with Eq.~\eqref{e3}.

We thus see that the gauge transformations of Eq.~\eqref{eq:35} are all closed. It is now a straightforward exercise to obtain the BRST transformations that leave the effective Lagrangian of Eq.~\eqref{eq:37} invariant. We identify $ ( \phi_{i} , \Phi_{I} )$ with $ (e_{\mu}^{a} , \omega_{\mu ab} )$, $(b_{i} , B_{I} )$ with $ (b_{\mu} , B_{ab} )$, the gauge transformations of Eq.~\eqref{eq:15} with these of Eq.~\eqref{eq:35}, the gauge functions $( \theta_{i} , \Theta_{I} )$ with $( \epsilon_{\mu} , \lambda_{ab} )$, the gauge condition of Eq.~\eqref{eq:16} with those of Eq.~\eqref{eq:36}, and the ghosts $(c_{i} , \bar{c}_{i} , C_{I} , \bar{C}_{I} )$ with $ (c_{\mu} , \bar{c}_{\mu} , C_{ab} , \bar{C}_{ab} )$. Now it is possible to use Eqs.~\eqref{eq:22}, \eqref{eq:23}, \eqref{eq:24}, \eqref{eq:28} to find the BRST transformations associated with $ \mathcal{L}_{\text{eff}} $ in Eq.~\eqref{eq:37}. 
Then, by using Eqs.~\eqref{eq:44}, \eqref{eq:46} and \eqref{eq:49}, one finds that Eq.~\eqref{eq:28} for the BRST transformations of the ghost fields $( c^{\mu} , C^{ab} )$ have the form (see Eqs.~\eqref{eq:B4}, \eqref{eq:B7}, \eqref{eq:B10} and \eqref{eq:B12} in Appendix B)
\begin{subequations}\label{eq:BRST}
    \begin{align} \label{eq:BRST1}
        \delta c^{\mu} ={}&- c^{\alpha} A_{\alpha \beta}^{\mu}  c^{\beta} \eta  = c^{\alpha} (\tensor{c}{^{\mu}_{; \alpha}} - \tensor{S}{_{\alpha \beta}^{\mu}}c^{\beta}) \eta , 
        \\ \intertext{and}
        \delta C^{ab} ={}& - C^{lm} B^{ab}_{lm pq}  C^{pq} \eta - c^{\mu} B_{\mu  lm}^{ab} C^{lm} \eta =\left( -\tensor{C}{^{a}_{k}} C^{kb}   + c^{\mu} \tensor{C}{^{ab}_{, \mu}} \right)\eta  .
    \end{align}
\end{subequations}
It may be verified that these transformations are nilpotent (see Eqs.~\eqref{eq:B15a} and \eqref{eq:B16a}). 

 Eqs.~\eqref{eq:49} and \eqref{eq:52} indicate that the BRST transformations are separated into two parts, namely, the diffeomorphism and local Lorentz transformations. 
Thus by setting $ \xi^{\sigma} = c^{\sigma} \eta $ and $ \lambda_{ab} = C_{ab} \eta $, we have that 
\begin{equation}\label{eq:424}
     \{ \delta^{\text{LT}} , \ \delta^{\text{Diff}} \} = 0,
\end{equation}
where $ \delta_{\text{Diff}} $, $ \delta_{ \text{LT} } $ generates respectively diffeomorphism and local Lorentz BRST transformations. We note that, generally, the full transformation in a Riemann-Cartan manifold is composed of an holonomic diffeomorphism transformation  and an anholonomic local Lorentz transformation:
\begin{equation}\label{eq:425}
    \delta^{\text{BRST} } = \delta^{\text{Diff}} + \delta^{\text{LT}},
\end{equation}
which is consistent with Eqs.~\eqref{eq:32}, \eqref{eq:BRST} (and other results shown in the Appendix B). 
Then, using Eq.~\eqref{eq:424} and the nilpotency of $ \delta^{\text{Diff}}$ and $ \delta^{\text{LT}} $, we find that 
\begin{equation}\label{eq:426}
    \left(\delta^{ \text{BRST}}\right)^{2} =0
\end{equation}
which shows that the full operator $ \delta^{\text{BRST}} $ in Eq.~\eqref{eq:425} generates a nilpotent BRST transformation.

\section{Slavnov-Taylor identities}\label{section:STID}

In order to derive the Slavnov-Taylor identities, we start from the generating functional $ Z$ of Green's functions, which is required to be gauge invariant. This may be written in a compact form in terms of path integrals as 
\begin{equation}\label{eq:5.1}
    Z = \int \mathop{\mathcal{D} e_{\mu}^{a}} \mathop{\mathcal{D} \omega_{\mu ab}} \mathop{\mathcal{D} \psi} \mathop{\mathcal{D} \bar{\psi}} \mathop{\mathcal{D} c_{\mu}} \mathop{\mathcal{D} \bar{c}_{\mu}} \mathop{\mathcal{D} C_{ab}} \mathop{\mathcal{D} \bar{C}_{ab}} \exp i \int \mathop{d^{4}x} \mathcal{L}_{\text{eff}},
\end{equation}
where $ \mathcal{L}_{\text{eff}} $ is the effective Lagrangian \eqref{eq:37}, which contains the $ \mathcal{L}_{\text{EC}} $ and $ \mathcal{L}_{\text{D}} $ Lagrangians as well as the gauge-fixing terms and the Faddeev-Popov ghosts. 
Following the work done in Refs. \cite{Taylor:1971ff, Slavnov:1972fg, Kluberg-Stern:1974iel}, it is convenient to introduce in $Z$ the sources $S_i$ that couple to the fields which are denoted generically by $ \varphi_{i} $. Moreover, one also introduces the sources $ U_i$ which couple to the BRST variations $ \delta \varphi_{i} $ of the fields. Thus, one gets 
\begin{equation}\label{eq:5.2}
    Z [ S_{i} , U_{i} ] = \int \prod_{j}^{} \mathop{\mathcal{D} \varphi_{j}} \exp{ i \int \mathop{d^{4}x} \mathcal{L}_{\text{tot}}} ,
\end{equation}
where 
\begin{equation}\label{eq:5.3}
    \mathcal{L}_{tot} = \mathcal{L}_{\text{eff}} + S_{i} \varphi_{i} + U_{i} \delta \varphi_{i} . 
\end{equation}

We note here that, in consequence of the nilpotency of the BRST operator, the coefficients of the sources $U_i$ are gauge invariant. Due to this property and since $ \mathcal{L}_{\text{eff}} $ is invariant under the BRST transformations, 
it follows that
\begin{equation}\label{eq:5.4}
    Z [ S_{i} , U_{i} ]  \simeq \int \prod_{j} \mathop{\mathcal{D} \varphi_{j} } \left[ 1+ i \int \mathop{d^{4}x} S_{i} \delta \varphi_{i} \right] \exp{i \int \mathop{d^{4}x} \mathcal{L}_{ \text{tot} }}, 
\end{equation}
Thus, the invariance of $Z$ requires that 
\begin{equation}\label{eq:5.5}
    \int \prod_{j }^{} \mathop{\mathcal{D} \varphi_{j}} \int \mathop{d^{4} x} S_{i} \delta \varphi_{i} \exp i \int \mathop{d^{4}x} \mathcal{L}_{\text{tot}}=0,
\end{equation}
where the BRST variations of the fields are given by Eqs.~\eqref{eq:24}, \eqref{eq:32} and \eqref{eq:BRST}. Since the quantities $ \delta \varphi_{i} $ are the coefficients of the sources $ U_{i} $, Eq.~\eqref{eq:5.5} implies that 
\begin{equation}\label{eq:5.6}
    \int \mathop{d^{4} x} S_{i} \frac{\delta_{l} Z[S_i, U_i] }{\delta U_{i}} =0,
\end{equation}
where the subscript $l$ ($r$) denotes left (right) differentiation. 
This equation contains only first order derivatives, which is a consequence of the introduction of the sources $U_i$ for the non-linear variations $ \delta \varphi_{i} $ of the fields. Putting $Z [S_i,U_i] = \exp i W [S_i, U_i]$, a similar equation holds for $W [S_i, U_i]$.

We can now convert this into a condition on the non-renormalized generating functional $ \Gamma $ of one-particle irreducible Green's functions. To this end, one makes a Legendre transform in the usual way, except that we do not transform the sources $U_i$: (For simplicity of notation, we omit the averaging symbols on the fields $\varphi_i$)
\begin{equation}\label{eq:5.7}
    W [ S_i, U_i] = \Gamma [ \varphi_i, U_i] + \int \mathop{d^{4} x} S_i \varphi_{i}.
\end{equation}
Then, it turns out that $ \Gamma $ satisfies the Zinn-Justin master equation \cite{Zinn-Justin:2011bia} 
\begin{equation}\label{eq:5.8}
    \int \mathop{d^{4}x} \frac{\delta_{r} \Gamma }{\delta \varphi_{i}} \frac{\delta_{l} \Gamma }{\delta U_{i} } =0 . 
\end{equation}
These identities reflect the gauge-invariance of the theory. For example, to lowest order, one may replace here $ \Gamma $ by the tree-level effective action $ \Gamma_{0} = \int \mathop{d^{4}x} \mathcal{L}_{\text{eff}} $ (see Eq.~\eqref{eq:37}), and $ \delta_{l} \Gamma / \delta U_{i} $ by $ \delta \varphi_{i} $, to get the equation 
\begin{equation}\label{eq:5.9}
    \int \mathop{d^{4}x} \frac{\delta_{r} \Gamma_{0} }{\delta \varphi_{i}} \delta \varphi_{i} =0
\end{equation}
which states that $ \Gamma_{0} $ is BRST invariant. 

Using the relations \eqref{eq:23}, \eqref{eq:24}, \eqref{eq:5.8} and the background field method, one can show that in the first order EC theory one gets, in terms of the functional $ \Gamma '$ obtained from $ \Gamma $ by subtracting the gauge-fixing terms, the equation 
\begin{equation}\label{eq:5.10}
    \int \mathop{d^{4} x} \left [ \frac{\delta_{r} \Gamma ' }{\delta q_{a \nu}} \frac{\delta_{l} \Gamma' }{\delta U^{a \nu}} 
+ \frac{\delta_{r} \Gamma ' }{\delta Q_{\nu ab}} \frac{\delta_{l} \Gamma' }{\delta U^{\nu ab }} + \frac{\delta_{r} \Gamma ' }{\delta c_{ \nu}} \frac{\delta_{l} \Gamma' }{\delta U^{\nu}}  
+ \frac{\delta_{r} \Gamma ' }{\delta C_{a b}} \frac{\delta_{l} \Gamma' }{\delta U^{ab}} 
+ \frac{\delta_{r} \Gamma ' }{\delta \psi } \frac{\delta_{l} \Gamma' }{\delta U_{\bar{\psi}} } 
+ \frac{\delta_{l} \Gamma' }{\delta U_{\psi} }\frac{\delta_{r} \Gamma ' }{\delta \bar{\psi} }  
    \right ] =0,
\end{equation}
where $ q_{a \nu} $ and $ Q_{\nu ab} $ are the quantum tetrad and spin connection fields (see Eqs.~\eqref{eq:31} and \eqref{eq:32}).

The last two terms in Eq.~\eqref{eq:5.10} arise due to the fact that under local Lorentz transformations, the Dirac field $ \psi $ transforms as (see Eq.~\eqref{e12} with $ \lambda_{ab} = C_{ab} \eta$) 
\begin{equation}\label{eq:5.11}
    \delta^{\text{LT}} \psi = -\frac{1}{2} C_{ab} \sigma^{ab} \psi \eta .
\end{equation}
Since $ \psi $ is a scalar under diffeormorphism transformations, it follows that the full BRST transformation of the Dirac field $ \psi $ is given by (setting $ \epsilon^{\mu} = c^{\mu} \eta $) 
\begin{equation}\label{eq:5.12}
    \delta \psi = \left (  c^{\mu} \partial_{\mu} - \frac{1}{2} C_{ab} \sigma^{ab}  \right ) \psi \eta ,
\end{equation}
which is consistent with Eq.~\eqref{eq:425}. 
Thus, by introducing sources $ U_{\psi } $ and $ U_{\bar{\psi}} $ coupled to the fermions as $ U_{\bar{\psi}} \delta \psi +  \delta \bar{\psi} U_{\psi}$, the Dirac field contribution to the Zinn-Justin equation may be written in the form shown in Eq.~\eqref{eq:5.10}.

Since this theory possesses two symmetries and the BRST transformations associated with the diffeormorphism and local Lorentz invariance, anticommute, Eq.~\eqref{eq:5.10} can be split into two separate equations. This could be achieved by introducing two different sources $ u_{i} $ and $ \mathfrak{u}_{i} $ corresponding to the two distinct BRST variations $ \delta^{\text{Diff}} $ and $ \delta^{\text{LT}} $, which would lead to 
\begin{subequations}\label{eq:5.14}
    \begin{align}
    \int \mathop{d^{4} x} \left [ 
        \frac{\delta_{r} \Gamma ' }{\delta q_{a \nu}} \frac{\delta_{l} \Gamma' }{\delta u^{a \nu}} 
+ \frac{\delta_{r} \Gamma ' }{\delta Q_{\nu ab}} \frac{\delta_{l} \Gamma' }{\delta u^{\nu ab }} + \frac{\delta_{l} \Gamma ' }{\delta c_{ \nu}} \frac{\delta_{l} \Gamma' }{\delta u^{\nu}}  
+ \frac{\delta_{r} \Gamma ' }{\delta C_{a b}} \frac{\delta_{l} \Gamma' }{\delta u^{ab}   } 
+ \frac{\delta_{r} \Gamma ' }{\delta \psi } \frac{\delta_{l} \Gamma' }{\delta u_{\bar{\psi}} } 
+ \frac{\delta_{l} \Gamma' }{\delta u_{\psi} } \frac{\delta_{r} \Gamma ' }{\delta \bar{\psi} }  
\right ] ={}&0, 
\\   \int \mathop{d^{4} x} \left [ 
    \frac{\delta_{r} \Gamma ' }{\delta q_{a \nu}} \frac{\delta_{l} \Gamma' }{\delta \mathfrak{u}^{a \nu}} 
+ \frac{\delta_{r} \Gamma ' }{\delta Q_{\nu ab}} \frac{\delta_{l} \Gamma' }{\delta \mathfrak{u}^{\nu ab }} 
+ \frac{\delta_{r} \Gamma ' }{\delta C_{a b}} \frac{\delta_{l} \Gamma' }{\delta \mathfrak{u}^{ab}} 
+ \frac{\delta_{r} \Gamma ' }{\delta \psi } \frac{\delta_{l} \Gamma' }{\delta \mathfrak{u}_{\bar{\psi}} } 
+ \frac{\delta_{l} \Gamma' }{\delta \mathfrak{u}_{\psi} } \frac{\delta_{r} \Gamma ' }{\delta \bar{\psi} }  
\right ] ={}&0.
\end{align}
\end{subequations}
The master equation \eqref{eq:5.10} can be obtained using that  
\begin{equation}\label{eq:5.13}
    \frac{\delta_{l} \Gamma ' }{\delta U_{i}} = \frac{\delta_{l} \Gamma ' }{\delta u_{i} } + \frac{\delta_{l} \Gamma ' }{\delta \mathfrak{u}_{i}},
\end{equation}
which follows from Eq.~\eqref{eq:425}.

We obtain the Slavnov-Taylor identities by taking functional derivatives of Eq.~\eqref{eq:5.10}  with respect to the fields, which relate the proper Green's functions of the theory.
The master equation \eqref{eq:5.10} is useful to implement a systematic renormalization of the EC theory in the first order form. (see also Refs. \cite{Frenkel:2017xvm, Lavrov:2019nuz}) 

\section{Discussion}\label{section:discussion}

We have examined the quantization of EC theory in the first order form, where the tetrad and the spin connection are treated as two independent gauge fields.
This formulation embodies two distinct gauge symmetries, namely, the diffeomorphism and the local Lorentz invariance. 
We have shown that the gauge algebra is closed, even in the presence of torsion, and established the BRST invariances of the effective Lagrangian \eqref{eq:37}. 

We found that the algebra of the BRST generators separates into two distinct classes: 
\begin{subequations} \label{eq:61}
\begin{align}\label{eq:61a}
    \{ \delta_{\xi}^{\text{Diff}} ,\ \delta_{\zeta}^{\text{Diff}} \} ={}& \delta_{\Omega}^{\text{Diff}}, \\ \label{eq:61b}
    \{ \delta_{\lambda}^{\text{LT}} ,\  \delta_{\rho}^{\text{LT}} \} ={}& \delta_{L}^{\text{LT}}, \\ \label{eq:61c}
    \{ \delta^{ \text{Diff}} , \ \delta^{\text{LT}} \} ={}& 0;
\end{align}
\end{subequations}
where $ \delta^{\text{Diff}} $ and $ \delta^{\text{LT}} $ are the BRST generators of the diffeomorphism and local Lorentz transformations. Here, the quantities $ \Omega $ and $L$ are defined, respectively, in Eqs.~\eqref{eq:defOmega} and \eqref{eq:46}. We note that in Eq.~\eqref{eq:61}, there appear anticommutators (instead of commutators) due to the grassmannian nature of the BRST transformations.

The relation \eqref{eq:61c} and the nilpotency of these operators ensure that the generator of full BRST transformations
\begin{equation}\label{eq:a64}
    \delta^{\text{BRST}} = \delta^{\text{Diff}} + \delta^{\text{LT} }
\end{equation}
is also a nilpotent operator. This condition is necessary for a consistent BRST quantization of the EC theory.

We have also derived the Slavnov-Taylor identities, which reflect the gauge invariances of this theory under BRST transformations. These identities are important for a recursive renormalization of the EC theory in the first order form. This requires the inclusion of all possible counterterms allowed by its gauge symmetries which should cancel, order by order in the loop expansion, all ultraviolet divergences. The background field method may allow to carry out such a renormalization in a gauge invariant manner. This is currently being examined. 

An interesting application of the first-order form of the EC action could be in supergravity \cite{VanNieuwenhuizen:1981ae, Freedman:2012zz}, which would involve considering a local supersymmetry gauge transformation in conjunction with those of Eq.~\eqref{eq:35}. 

The presence of a BRST symmetry generally ensures that a theory is consistent with unitarity \cite{Kugo:1979gm}. One should note though that if one were to use background field quantization, then terms linear in the quantum field must be omitted \cite{Abbott:1980hw, Abbott:1981ke}. Since this breaks the BRST invariance, special care must be exercised to use this invariance to prove unitary \cite{Frenkel:2018xup}. It is expected that through the use of a Lagrange multiplier field to ensure that the classical equations of motion are satisfied, all radiative corrections beyond one loop order can be eliminated yielding a unitary and renormalizable theory \cite{Brandt:2020gms}.
In this alternative formalism, the renormalizability of the EC theory in the first order form may be proven, even when coupled to matter fields \cite{Brandt:2021qgh}.

\begin{acknowledgments}
D.\ G.\ C\@. M\@. thanks Ann Aksoy for enlightening conversations and
J. Peters and T. Thrasher for their assistance.
F.\ T.\ B\@., J.\ F\@. and S.\ M.-F\@. thank CNPq (Brazil) for financial support. 
\end{acknowledgments}

\appendix

\section{Closure of the gauge algebra in the EC theory}

In Eq.~\eqref{eq:39}, it is shown that the gauge algebra of transformations of the form in Eq.~\eqref{eq:38} is closed. In this appendix, we will show the derivation of Eq.~\eqref{eq:38} in detail. Before that, we will introduce some notation and derive some useful identities used throughout this appendix.

\subsection{Notation and useful identities}

Let us consider the four-vectors $ V_{\mu} $ and $ V^{\mu} $. 
A semicolon denotes a covariant derivative:
\begin{equation}\label{eq:A1}
    \begin{split}
        V_{\alpha ; \mu} ={}&V_{\alpha , \mu} - \tensor{\Gamma}{_{\mu \alpha}^{\beta}}V_{\beta}, \\ 
        \tensor{V}{^{\alpha}_{; \mu}} ={}&V_{\alpha , \mu} + \tensor{\Gamma}{_{\mu \beta}^{\alpha} }V^{\beta};\\ 
\end{split} 
\end{equation}
where $ \tensor{\Gamma}{_{\mu \nu}^{\lambda} } $ is the affine connection. 
A comma denotes the partial derivative, $ V_{\alpha , \mu} \equiv  \partial_{\mu} V_{\alpha} $. 

The antisymmetrization of indices is denoted by brackets, for example, 
\begin{equation}\label{eq:Riemann}
    R_{[ \mu \sigma \lambda ] \alpha} \equiv  \frac{1}{3!} \left ( R_{\mu \sigma \lambda \alpha} + R_{\sigma \lambda \mu \alpha} + R_{\lambda \mu \sigma \alpha} - R_{\sigma \mu \lambda \alpha } - R_{\lambda \sigma \mu \alpha } - R_{\mu \lambda \sigma \alpha }\right ). 
\end{equation}

If we take $ R_{\mu \sigma \lambda \alpha} $ to represent the Riemann curvature tensor defined by
\begin{equation}\label{eq:F1c} 
    \tensor{R}{_{\mu \nu \beta }^{\alpha} } =  \tensor{\Gamma}{_{\nu \beta}^{\alpha}_{, \mu} } -  \tensor{\Gamma}{_{\mu \beta}^{\alpha}_{,\nu} } + 
    \tensor{\Gamma}{_{\mu \lambda}^{\alpha}} \tensor{\Gamma}{_{\nu \beta}^{\lambda}}
    -
    \tensor{\Gamma}{_{\nu \lambda}^{\alpha}} \tensor{\Gamma}{_{\mu \beta}^{\lambda}},
\end{equation}
which is antisymmetric in the first two indices:
\begin{equation}\label{eq:syofR}
    R_{\mu \nu  \alpha \beta}  = - R_{\nu \mu \alpha \beta} .
\end{equation}
In the EC theory the metricity condition $ g_{\mu \nu ; \lambda} =0$ is satisfied, thus we also have that $ R_{\mu \nu \alpha \beta} = -R_{\mu \nu \beta \alpha}$. 
Using Eq.~\eqref{eq:syofR} in the Eq.~\eqref{eq:Riemann} reads
\begin{equation}\label{eq:RiemannTotal}
    R_{[ \mu \sigma \lambda ] \alpha} = \frac{1}{3} \left ( R_{\mu \sigma \lambda \alpha} + R_{\sigma \lambda \mu \alpha} + R_{\lambda \mu \sigma \alpha}\right ).
\end{equation}

In a general space-time, we have that 
\begin{subequations}
\label{eq:F1b}
\begin{align}\label{eq:F1ba}
V_{\alpha;\mu\nu} -V_{\alpha ;\nu\mu} 
    ={}& 
    -\tensor{R}{_{ \nu \mu \alpha}^{\beta }} V_{\beta} + 2 \tensor{S}{_{\mu \nu}^{\lambda}}  V_{\alpha; \lambda }, \\ \label{eq:F1bb}
    \tensor{V}{^{\alpha}_{; \mu \nu}} - \tensor{V}{^{\alpha}_{;\nu \mu}}
={}& \tensor{R}{_{\nu \mu  \beta}^{\alpha }} V^{\beta} +2 \tensor{S}{_{\mu \nu}^{\lambda}}  \tensor{V}{^{\alpha}_{; \lambda }}.
\end{align}
\end{subequations}
 
\begin{proof}[Derivation of Eq.~\eqref{eq:F1ba}]
    Using the definition of the covariant derivative in Eq.~\eqref{eq:A1}, we obtain that 
    \begin{equation}\label{eq:dd1}
        \begin{split}
            V_{\alpha ; \mu \nu } ={}& (V_{\alpha , \mu} - \tensor{\Gamma}{_{\mu \alpha}^{\lambda}} V_{\lambda} )_{, \nu} - \tensor{\Gamma}{_{ \nu \alpha}^{\lambda} }
        (V_{\lambda , \mu} - \tensor{\Gamma}{_{\mu \lambda }^{\beta }} V_{\beta } )
- \tensor{\Gamma}{_{ \nu \mu}^{\lambda} }
        (V_{\alpha , \lambda} - \tensor{\Gamma}{_{ \lambda \alpha }^{\beta }} V_{ \beta } ) \\
            ={}& V_{\alpha , \mu \nu} - \tensor{\Gamma}{_{\mu \alpha}^{\lambda}_{, \nu}}V_{\lambda} - \tensor{\Gamma}{_{\mu \alpha}^{\lambda}}V_{\lambda , \nu} 
- \tensor{{\Gamma}}{_{ \nu \alpha}^{\lambda} }
        (V_{\lambda , \mu} - \tensor{{\Gamma}}{_{\mu \lambda }^{\beta }} V_{\beta } )
- \tensor{{\Gamma}}{_{ \nu \mu}^{\lambda} }
        (V_{\alpha , \lambda} - \tensor{{\Gamma}}{_{ \lambda \alpha }^{\beta }} V_{ \beta } ).
    \end{split} 
    \end{equation}
    Then, 
    \begin{equation}\label{eq:dd2}
        \begin{split}
            V_{\alpha ; \mu \nu} - V_{\alpha ; \nu \mu} ={}& -( \tensor{\Gamma}{_{\mu \alpha}^{\beta}_{,\nu}} - 
        \tensor{\Gamma}{_{\nu \alpha}^{\beta}_{,\mu}}
        -\tensor{\Gamma}{_{\nu \alpha}^{ \lambda}} \tensor{\Gamma}{_{\mu \lambda}^{\beta}} +        \tensor{\Gamma}{_{\mu \alpha}^{ \lambda}} \tensor{\Gamma}{_{\nu \lambda}^{\beta}}) V_{\beta} + 2 \tensor{S}{_{\mu \nu}^{\lambda} }  V_{\alpha ; \lambda }  
        \\
        ={}&
        - \tensor{R}{_{\nu \mu\alpha}^{\beta} } V_{\beta} + 2 \tensor{S}{_{\mu \nu}^{\lambda}}  V_{\alpha ; \lambda }.
    \end{split}
    \end{equation}
    By Eq.~\eqref{eq:syofR}, we see that Eq.~\eqref{eq:dd2} is equal to Eq.~\eqref{eq:F1ba}.
    The demonstration of Eq.~\eqref{eq:F1bb} is similar. 
\end{proof}

The Bianchi symmetry identity, in a general space-time, is given by \cite{Penrose:1983mf}
\begin{equation}\label{eq:BianchiwithTorsion}
    \tensor{R}{_{[ \alpha \beta \gamma]}^{\delta}} = 2 \tensor{S}{_{[\beta \gamma}^{\delta}_{;\alpha ]}} 
    -4\tensor{S}{_{[ \alpha \beta}^{\rho}} \tensor{S}{_{\gamma ]\rho}^{\delta} }.
\end{equation}

\subsection{Covariant gauge transformations}

The gauge transformation of the form of Eq.~\eqref{e8} is not manifestly covariant. However, using that 
\begin{equation}\label{eq:p2}
    V_{\mu ; \lambda} = V_{\mu , \lambda} - \tensor{\Gamma}{_{\lambda \mu}^{\alpha}}V_{\alpha}  
\end{equation}
and 
\begin{equation}\label{eq:p2a}
    \tensor{\epsilon}{^{\lambda}_{; \mu}} = \tensor{\epsilon}{^{\lambda}_{, \mu}} + \tensor{\Gamma}{_{\mu \alpha}^{\lambda}} \epsilon^{\alpha},
\end{equation}
we get that 
\begin{equation}\label{eq:p3}
    -\epsilon^{\lambda} V_{\mu ; \lambda} -V_{\lambda} \tensor{\epsilon}{^{\lambda}_{; \mu}} =
    -\epsilon^\lambda V_{\mu , \lambda}  - V_\lambda \tensor{\epsilon}{^{\lambda}_{, \mu}} 
+ \epsilon^{\lambda } \tensor{\Gamma}{_{\lambda \mu}^{\alpha}} V_{\alpha} - V_{\lambda}  \tensor{\Gamma}{_{\mu \alpha}^{\lambda}} 
\epsilon^{\alpha}
\end{equation}
yielding 
\begin{equation}\label{eq:p4}
    -\epsilon^\lambda V_{\mu , \lambda}  - V_\lambda \tensor{\epsilon}{^{\lambda}_{, \mu}} 
   =
-\epsilon^{\lambda} V_{\mu ; \lambda} -V_{\lambda} \tensor{\epsilon}{^{\lambda}_{; \mu}} - \epsilon^{\lambda} V_{\alpha} ( \tensor{\Gamma}{_{\lambda \mu}^{\alpha}} - \tensor{\Gamma}{_{\mu \lambda}^{\alpha}} ).
\end{equation}
Hence, Eq.~\eqref{e8a} can be written in a covariant way as
\begin{equation}\label{eq:covariantp1}
        \delta_\epsilon V_{\mu} (x)= 
    -\epsilon^{\lambda} V_{\mu ; \lambda} -V_{\lambda} \tensor{\epsilon}{^{\lambda}_{; \mu}} +2 \tensor{S}{_{\mu \lambda }^{\alpha} }   V_{\alpha}\epsilon^{\lambda}.
\end{equation}
Similarly, we also have  
\begin{equation}\label{eq:covariantp2}
        \delta_\epsilon V^{\mu} (x) =
        - \epsilon^{\lambda} \tensor{V}{^{\mu}_{; \lambda}} + V^{\lambda} \tensor{\epsilon}{^{\mu}_{; \lambda}}+ 2 \tensor{S}{_{\lambda \alpha}^{\mu}}V^{\alpha} \epsilon^{\lambda}
\end{equation}
which is the covariant formulation of Eq.~\eqref{e8b}.  Eq.~\eqref{eq:35} is a special case of the identity \eqref{eq:covariantp1} in the context of the background field method. 

The Eq.~\eqref{eq:covariantp1} can be straightforwardly generalized for a tensor field $T_{\mu \nu}$:
\begin{equation}\label{eq:covarianttensor1}
    \delta_\epsilon T_{\mu \nu } (x) =
    -\epsilon^{\lambda} T_{\mu \nu  ; \lambda} -T_{\mu \lambda  } \tensor{\epsilon}{^{\lambda}_{; \nu}}-T_{\lambda \nu } \tensor{\epsilon}{^{\lambda}_{; \mu}}  +2 \tensor{S}{_{\mu \lambda }^{\alpha} }   T_{\alpha \nu }\epsilon^{\lambda} 
    +2 \tensor{S}{_{\nu \lambda }^{\alpha} }   T_{\mu \alpha  }\epsilon^{\lambda} .
\end{equation}

\subsection{Derivation of Eqs.~\eqref{eq:39} and \eqref{eq:44}} 

We will obtain the second variation of $ q_{\mu}^{a} $ in Eq.~\eqref{eq:38} in two parts. First, let us consider the variation of
\begin{equation} \label{eq:defW}
    W_{\mu}^{a}[ \zeta ]= - \zeta^{\lambda} e_{\mu ; {\lambda }}^{a} - e_{\lambda}^{a} \tensor{\zeta}{^{\lambda}_{; {\mu}}}
\end{equation}
which does not explicitly depend on the torsion. By using Eq.~\eqref{eq:covariantp1}, we find  
\begin{equation}\label{eq:secondVar}
    \begin{split}
        \delta_{\xi} W_{\mu}^{a} [ \zeta ] ={}& -\xi^{\sigma} W_{\mu ; \sigma}^{a} [ \zeta ]- W_{\sigma}^{a} [\zeta ] \tensor{\xi}{^{\sigma}_{; \mu}} + 2 \tensor{S}{ _{\mu \nu}^{\sigma}} W_{\sigma}^{a} [ \zeta ]\xi^{\nu} 
    \\
        ={}& 
        \xi^{\sigma} (  \tensor{\zeta}{^{\lambda}_{; \sigma}} e_{\mu ; \lambda}^{a} + \zeta^{\lambda} e_{\mu ; \lambda \sigma}^{a} + e_{\lambda ; \sigma}^{a} \tensor{\zeta}{^{\lambda}_{; \mu}}+ e_{\lambda}^{a} \tensor{\zeta}{^{\lambda}_{; \mu \sigma}}) 
        + (\zeta^{\lambda} e_{\sigma ; {\lambda }}^{a} + e_{\lambda}^{a} \tensor{\zeta}{^{\lambda}_{; {\sigma}}}) \tensor{\xi}{^{\sigma}_{; \mu}} 
        - 2 \tensor{S}{_{\mu \nu}^{\sigma}} (\zeta^{\lambda} e_{\sigma ; {\lambda }}^{a} + e_{\lambda}^{a} \tensor{\zeta}{^{\lambda}_{; {\sigma}}}) \xi^{\nu}.
    \end{split} 
\end{equation}
Thus, the commutator reads
\begin{equation}\label{eq:qq1}
    \begin{split}
        \delta_{\xi} W_{\mu}^{a}  [ \zeta ]- \delta_{\zeta}  W_{\mu}^{a} [ \xi ] 
        ={}&
        \xi^{\sigma} (  \tensor{\zeta}{^{\lambda}_{; \sigma}} e_{\mu ; \lambda}^{a} + \zeta^{\lambda} e_{\mu ; \lambda \sigma}^{a} + e_{\lambda ; \sigma}^{a} \tensor{\zeta}{^{\lambda}_{; \mu}}+ e_{\lambda}^{a} \tensor{\zeta}{^{\lambda}_{; \mu \sigma}}) 
        -
 \zeta^{\sigma} (  \tensor{\xi}{^{\lambda}_{; \sigma}} e_{\mu ; \lambda}^{a} + \xi^{\lambda} e_{\mu ; \lambda \sigma}^{a} + e_{\lambda ; \sigma}^{a} \tensor{\xi}{^{\lambda}_{; \mu}}+ e_{\lambda}^{a} \tensor{\xi}{^{\lambda}_{; \mu \sigma}}) 
        \\ & \quad  
        + (\zeta^{\lambda} e_{\sigma ; {\lambda }}^{a} + e_{\lambda}^{a} \tensor{\zeta}{^{\lambda}_{; {\sigma}}}) \tensor{\xi}{^{\sigma}_{; \mu}}  
        -
        (\xi^{\lambda} e_{\sigma ; {\lambda }}^{a} + e_{\lambda}^{a} \tensor{\xi}{^{\lambda}_{; {\sigma}}}) \tensor{\zeta}{^{\sigma}_{; \mu}}  
        \\ & \quad  + 2 \tensor{S}{_{\mu \nu}^{\sigma}} (\xi^{\lambda} e_{\sigma ; {\lambda }}^{a} +e_{\lambda}^{a} \tensor{\xi}{^{\lambda}_{; {\sigma}}}) \zeta^{\nu}- 2 \tensor{S}{_{\mu \nu}^{\sigma}} (\zeta^{\lambda} e_{\sigma ; {\lambda }}^{a} + e_{\lambda}^{a} \tensor{\zeta}{^{\lambda}_{; \sigma}}) \xi^{\nu} 
        \\
        ={}& 
        -\tilde{\Omega}^{\lambda}  e_{\mu ; \lambda}^{a} -e_{\lambda}^{a} \tensor{\tilde{\Omega}}{^{\lambda}_{; \mu} }  + (e^{a}_{\mu ; \lambda \sigma} - e^{a}_{\mu ; \sigma \lambda}) \xi^{\sigma} \zeta^{\lambda}  
        + e_{\lambda}^{a} \left( 
            - \zeta^{\sigma} \tensor{\xi}{^{\lambda}_{; \mu \sigma}}
            + \zeta^{\sigma} \tensor{\xi}{^{\lambda}_{; \sigma \mu}}
            +\xi^{\sigma} \tensor{\zeta}{^{\lambda}_{; \mu \sigma}}
            - \xi^{\sigma} \tensor{\zeta}{^{\lambda}_{; \sigma \mu}}
        \right)
        \\ & \quad 
        + 2 \tensor{S}{_{\mu \nu}^{\sigma}} e_{\sigma ; \lambda}^{a} (\xi^{\lambda} \zeta^{\nu} - \zeta^{\lambda} \xi^{\nu} ) 
        + 2\tensor{S}{_{\mu \nu}^{\sigma}}e_{\lambda}^{a} ( \tensor{\xi}{^{\lambda}_{;\sigma}} \zeta^{\nu} - \tensor{\zeta}{^{\lambda}_{; \sigma} } \xi^{\nu} ),
    \end{split} 
\end{equation}
where  $  \tilde{\Omega}^{\lambda} \equiv -(\xi^{\sigma}\tensor{\zeta}{^{\lambda}_{;\sigma}} -\zeta^{\sigma}\tensor{\xi}{^{\lambda}_{; \sigma}} )  $.

Now, we obtain the variation of the last term of Eq.~\eqref{eq:38}:
\begin{equation}\label{eq:q12}
    \delta_{\xi} Y_{\mu}^{a} [\zeta ] \equiv  \delta_{\xi}\left( 2 \tensor{S}{_{\mu \nu}^{\beta}} e_{\beta}^a \zeta^{\nu } \right)
    =
-2 \xi^{\lambda}\left (  S_{\mu \nu}^{a} \zeta^{\nu} \right )_{; \lambda} -2 S_{ \lambda \nu}^{a} \zeta^{\nu} \tensor{\xi}{^{\lambda}_{; \mu}} + 4 \tensor{S}{_{\mu \lambda}^{\beta} } S_{ \beta \nu}^{a} \zeta^{\nu} \xi^{\lambda},
\end{equation}
where $    S_{\mu \nu}^{a} \equiv \tensor{S}{_{\mu \nu}^{\beta}} e_{\beta}^a  $.
The commutator for this part is given by
\begin{equation}\label{eq:q3}
    \begin{split}
        \delta_{\xi} Y_{\mu}^{a} [\zeta] - \delta_{\zeta}  Y_{\mu}^{a} [ \xi ] ={}& 
\delta_{\xi}\left( 2 \tensor{S}{_{\mu \nu}^{\beta}} e_{\beta}^a \zeta^{\nu } \right) - 
\delta_{\zeta}\left( 2 \tensor{S}{_{\mu \nu}^{\beta}} e_{\beta}^a \xi^{\nu } \right)
\\
        ={}&
        -2 S_{\mu \nu}^{a} ( \xi^{\lambda} \tensor{\zeta}{^{\nu}_{;\lambda}}- \zeta^{\lambda} \tensor{\xi}{^{\nu}_{;\lambda} } ) - 2S_{\mu \nu ; \lambda}^{a} ( \xi^{\lambda} \zeta^{\nu} - \zeta^{\lambda} \xi^{\nu} ) 
        \\ & \quad - 2 S_{\lambda \nu}^{a} ( \zeta^{\nu} \tensor{\xi}{^{\lambda}_{; \mu}}- \xi^{\nu} \tensor{\zeta}{^{\lambda}_{; \mu}}) + 4\tensor{S}{_{\mu \lambda}^{\beta} } S_{ \beta \nu}^{a} ( \zeta^{\nu} \xi^{\lambda} - \xi^{\nu} \zeta^{\lambda} ). 
        \\
        ={}&
        2 S_{\mu \nu}^{a} \tilde{\Omega}^{\nu}  - 2S_{\mu \nu ; \lambda}^{a} ( \xi^{\lambda} \zeta^{\nu} - \zeta^{\lambda} \xi^{\nu} ) - 2 S_{\lambda \nu}^{a} ( \zeta^{\nu} \tensor{\xi}{^{\lambda}_{; \mu}}- \xi^{\nu} \tensor{\zeta}{^{\lambda}_{; \mu}}) + 4\tensor{S}{_{\mu \lambda}^{\beta} } S_{ \beta \nu}^{a} ( \zeta^{\nu} \xi^{\lambda} - \xi^{\nu} \zeta^{\lambda} ). 
    \end{split}
\end{equation}

We obtain the total contribution to the commutator $ 
        \left ( \delta_{\xi} \delta_{\zeta} - \delta_{\zeta} \delta_{\xi}\right ) q_{\mu}^{a} $ by summing Eqs.~\eqref{eq:qq1} and \eqref{eq:q3}:
\begin{equation}\label{eq:39REVISED}
    \begin{split}
        \left ( \delta_{\xi} \delta_{\zeta} - \delta_{\zeta} \delta_{\xi}\right ) q_{\mu}^{a} ={}& 
-\tilde{\Omega}^{\lambda}  e_{\mu ; \lambda}^{a} -e_{\lambda}^{a} \tensor{\tilde{\Omega}}{^{\lambda}_{; \mu} }  
+          2 \tensor{S}{_{\mu \nu}^{\beta}} e_{\beta}^{a} \tilde{\Omega}^{\nu} \\ 
                                                                                                            & \quad + (e^{a}_{\mu ; \lambda \sigma} - e^{a}_{\mu ; \sigma \lambda}) \xi^{\sigma} \zeta^{\lambda}  
        + e_{\lambda}^{a} \left( 
            - \zeta^{\sigma} \tensor{\xi}{^{\lambda}_{; \mu \sigma}}
            + \zeta^{\sigma} \tensor{\xi}{^{\lambda}_{; \sigma \mu}}
            +\xi^{\sigma} \tensor{\zeta}{^{\lambda}_{; \mu \sigma}}
            - \xi^{\sigma} \tensor{\zeta}{^{\lambda}_{; \sigma \mu}}
    \right) \\ & \quad \quad   
+ (4\tensor{S}{_{\mu \lambda}^{\beta} } \tensor{S}{_{ \beta \nu}^{\alpha }} - 2\tensor{S}{_{\mu \nu }^{\alpha}_{; \lambda} }e^{a}_{\alpha } )(\xi^{\lambda} \zeta^{\nu} - \zeta^{\lambda} \xi^{\nu} ) 
            \\ & \quad \quad  
    - 2 S_{\lambda \nu}^{a} ( \zeta^{\nu} \tensor{\xi}{^{\lambda}_{; \mu}}- \xi^{\nu} \tensor{\zeta}{^{\lambda}_{; \mu}})  
        + 2\tensor{S}{_{\mu \nu}^{\sigma}}e_{\lambda}^{a} ( \tensor{\xi}{^{\lambda}_{;\sigma}} \zeta^{\nu} - \tensor{\zeta}{^{\lambda}_{; \sigma} } \xi^{\nu} ).
\end{split}
\end{equation}

Using the anti-symmetry of the torsion $ S_{\mu \nu \alpha} = - S_{\nu \mu \alpha } $, we can simplify the last three terms to
\begin{equation}\label{eq:zero0}
    -2 \xi^{\sigma} \zeta^{\lambda} e^{a}_{\alpha}( -2\tensor{S}{_{\mu [\sigma }^{\alpha}_{; \lambda]} } +4\tensor{S}{_{\mu [\sigma}^{\beta} } \tensor{S}{_{ \lambda ] \beta }^{\alpha }} ) 
    - 2 \tensor{S}{_{\lambda \nu}^{ \beta}} e_{\beta}^{a}   
    ( \zeta^{\nu} \tensor{\xi}{^{\lambda}_{; \mu}}- \xi^{\nu} \tensor{\zeta}{^{\lambda}_{; \mu}})  
        + 2\tensor{S}{_{\mu \sigma}^{\lambda}}e_{\beta}^{a} ( \tensor{\xi}{^{\beta}_{;\lambda}} \zeta^{\sigma} - \tensor{\zeta}{^{\beta}_{; \lambda} } \xi^{\sigma} ).
\end{equation}
Besides that, using Eq.~\eqref{eq:F1b} in 
\begin{equation}\label{eq:defZ}
        (e^{a}_{\mu ; \lambda \sigma} - e^{a}_{\mu ; \sigma \lambda}) \xi^{\sigma} \zeta^{\lambda}  
        + e_{\lambda}^{a} \left( 
            - \zeta^{\sigma} \tensor{\xi}{^{\lambda}_{; \mu \sigma}}
            + \zeta^{\sigma} \tensor{\xi}{^{\lambda}_{; \sigma \mu}}
            +\xi^{\sigma} \tensor{\zeta}{^{\lambda}_{; \mu \sigma}}
            - \xi^{\sigma} \tensor{\zeta}{^{\lambda}_{; \sigma \mu}}
    \right)
\end{equation}
yields
    \begin{equation}\label{eq:ee1}
        \xi^{\sigma} \zeta^{\lambda} e^{a \, \alpha} \left ( -R_{ \sigma \lambda  \mu \alpha} +R_{ \mu  \lambda  \sigma \alpha}+R_{\sigma \mu \lambda \alpha }   \right ) +2 \tensor{S}{_{\lambda \sigma }^{\beta}} \xi^{\sigma} \zeta^{\lambda} e_{\mu ; \beta }^{a}  + 2 e_{\lambda}^{a}  \tensor{S}{_{\sigma \mu}^{\beta}} (\tensor{\xi}{^{\lambda}_{; \beta}} \zeta^{\sigma } - \xi^{\sigma} \tensor{\zeta}{^{\lambda}_{; \beta}} ).
    \end{equation}
    By Eqs.~\eqref{eq:syofR} and \eqref{eq:RiemannTotal}, we see that
    \begin{equation}\label{eq:ee2}
  -R_{ \sigma \lambda  \mu \alpha} +R_{\mu \lambda \sigma \alpha } +R_{\sigma \mu \lambda \alpha } =-3 R_{[ \mu \sigma \lambda  ] \alpha}.
    \end{equation}
    Therefore, we can use the Bianchi symmetry identity \eqref{eq:BianchiwithTorsion} leading to
\begin{equation} \label{eq:F6bgen}
3 \xi^{\sigma} \zeta^{\lambda}e^{a}_{\alpha}  (-2  \tensor{S}{_{[\mu \sigma }^{\alpha}_{; \lambda ]} } +4 \tensor{S}{_{[ \mu \sigma}^{\beta}} \tensor{S}{_{\lambda]\beta}^\alpha}  ) 
-2 \tensor{S}{_{\sigma \lambda}^{\beta}} e_{\mu ; \beta}^{a} \zeta^{\lambda} \xi^{\sigma} 
        -2 \tensor{S}{_{\mu \sigma}^{\lambda}} e_{\beta}^{a} ( \zeta^{\sigma} \tensor{\xi}{^{\beta}_{; \lambda}}- \xi^{\sigma} \tensor{\zeta}{^{\beta}_{; \lambda }}).
\end{equation}
    in which we have used that $ S_{\mu \nu \alpha} = - S_{\nu \mu \alpha} $.

By inserting the results in Eqs.~\eqref{eq:zero0} and~\eqref{eq:F6bgen} into Eq.~\eqref{eq:39REVISED}, we obtain 
\begin{equation}\label{eq:38NEW}
\begin{split}
        \left ( \delta_{\xi} \delta_{\zeta} - \delta_{\zeta} \delta_{\xi}\right ) q_{\mu}^{a} ={}& 
-\tilde{\Omega}^{\lambda}  e_{\mu ; \lambda}^{a} -e_{\lambda}^{a} \tensor{\tilde{\Omega}}{^{\lambda}_{; \mu} }  
        +          2 \tensor{S}{_{\mu \nu}^{\beta}} e_{\beta}^{a} \tilde{\Omega}^{\nu} +  
\xi^{\sigma} \zeta^{\lambda} e_{\alpha}^{a} ( -2\tensor{S}{_{\sigma \lambda}^{\alpha}_{; \mu}} + 4 \tensor{S}{_{\sigma \lambda}^{\beta} } \tensor{S}{_{\mu \beta}^{\alpha}})
\\ & \quad   
-2 \tensor{S}{_{\sigma \lambda}^{\beta}} e_{\mu ; \beta}^{a} \zeta^{\lambda} \xi^{\sigma} 
- 2 \tensor{S}{_{\lambda \nu}^{ \beta}} e_{\beta}^{a}   
    ( \zeta^{\nu} \tensor{\xi}{^{\lambda}_{; \mu}}- \xi^{\nu} \tensor{\zeta}{^{\lambda}_{; \mu}})
\end{split}
\end{equation}
which is equal to Eq.~\eqref{eq:39}. 

Now, if we define 
\begin{equation}\label{eq:defnewpar0} 
    \Omega^{\beta} = \tilde{\Omega}^{\beta} + 2 \tensor{S}{_{\sigma \lambda}^{\beta} } \xi^{\sigma} \zeta^{\lambda},
\end{equation} we see that 
\begin{equation}\label{eq:defnewpar}
\tensor{\Omega}{^{\beta}_{; \mu}} = \tensor{\tilde{\Omega}}{^{\beta}_{; \mu}}+2 \tensor{S}{_{\sigma \lambda}^{\beta}_{; \mu}} \xi^{\sigma} \zeta^{\lambda} +2 \tensor{S}{_{\sigma \lambda}^{\beta}} ( \tensor{\xi}{^{\sigma}_{; \mu}} \zeta^{\lambda} - \tensor{\zeta}{^{\sigma}_{; \mu}} \xi^{\lambda}).
\end{equation}
Then, Eq.~\eqref{eq:38NEW} reads 
\begin{equation}\label{eq:44NEW}
        \left ( \delta_{\xi} \delta_{\zeta} - \delta_{\zeta} \delta_{\xi}\right ) q_{\mu}^{a} =
        -\Omega^{\lambda}  e_{\mu ; \lambda}^{a} -e_{\lambda}^{a} \tensor{\Omega}{^{\lambda}_{; \mu} }  
+          2 \tensor{S}{_{\mu \nu}^{\beta}} e_{\beta}^{a} \Omega^{\nu}.
\end{equation}

Note that, the gauge parameter $ \Omega^{\lambda} = \delta_{\xi} \zeta^{\lambda} = - \xi^{\sigma} \tensor{\zeta}{^{\lambda}_{, \sigma}}+ \zeta^{\sigma} \tensor{\xi}{^{\lambda}_{, \sigma}} $. By using Eq.~\eqref{eq:p3} we can rewrite Eq.~\eqref{eq:44NEW} as 
\begin{equation}\label{eq:44NEWind}
        \left ( \delta_{\xi} \delta_{\zeta} - \delta_{\zeta} \delta_{\xi}\right ) q_{\mu}^{a} =
        -\Omega^{\lambda}  e_{\mu , \lambda}^{a} -e_{\lambda}^{a} \tensor{\Omega}{^{\lambda}_{, \mu} }.
\end{equation}
In order to obtain Eq.~\eqref{eq:44NEWind} in a background covariant way, one can use Eq.~\eqref{eq:p3} with $ \tensor{\bar{\Gamma}}{_{\mu \nu}^{\lambda}}$ replacing $ \tensor{\Gamma}{_{\mu \nu}^{\lambda}}$ which yields Eq.~\eqref{eq:44}.

Since Eq.~\eqref{eq:44NEW} holds for any four-vector $ V_{\mu} $, it can also be generalized for any tensor. For simplicity, let us consider a tensor like $ T_{\mu \nu}^{ab} = V_{\mu}^{a} V_{\nu}^{a} $, which under diffeomorphisms transforms as in Eq.~\eqref{eq:covarianttensor1}. 
A first order transformation with a parameter $ \zeta $ of $ V_{\mu}^{a} V_{\nu}^{b} $ yields 
\begin{equation}\label{eq:16B}
    \delta_{\zeta} T_{\mu \nu}^{ab} =  V_{\nu}^{b}\delta_{\zeta} V_{\mu}^{a} + V_{\mu}^{a} \delta_{\zeta} V_{\nu}^{b}. 
\end{equation}
A second order variation (with parameter $ \xi $) gives: 
\begin{equation}\label{eq:16C}
    \delta_{\xi} \delta_\zeta T_{\mu \nu}^{ab} = V^{b}_{\nu} \delta_{\xi} \delta_{\zeta} V_{\mu}^{a} + ( \delta_{\zeta} V_{\mu}^{a} )(\delta_{\xi} V_{\nu}^{b} ) +  ( \delta_{\xi} V_{\mu}^{a} )( \delta_{\zeta} V_{\nu}^{b} ) + V_{\mu}^{a} \delta_{\xi} \delta_{\zeta} V_{\nu}^{b}.
\end{equation}
Subtracting from this a part with $ \xi \leftrightarrow \zeta $, leads to several cancellations, so we get
\begin{equation}\label{eq:16D}
    ( \delta_{\xi} \delta_{\zeta} - \delta_{\zeta} \delta_{\xi} ) 
    T_{\mu \nu}^{ab} = V_{\nu}^{b} 
    ( \delta_{\xi} \delta_{\zeta} - \delta_{\zeta} \delta_{\xi} ) V_{\mu}^{a} + V_{\mu}^{a}  
    ( \delta_{\xi} \delta_{\zeta} - \delta_{\zeta} \delta_{\xi} ) V_{\nu}^{b}. 
\end{equation}
Using the fact that 
\begin{equation}\label{eq:44general}
    \left ( \delta_{\xi} \delta_{\zeta} - \delta_{\zeta} \delta_{\xi}\right ) V_{\mu}^{a} =
        -\Omega^{\lambda}  V_{\mu ; \lambda}^{a} -V_{\lambda}^{a} \tensor{\Omega}{^{\lambda}_{; \mu} }  
+          2 \tensor{S}{_{\mu \lambda}^{\beta}} V_{\beta}^{a} \Omega^{\lambda},
\end{equation}
we obtain from Eq.~\eqref{eq:16D}: 
\begin{equation}\label{eq:16F}
    ( \delta_{\xi} \delta_{\zeta} - \delta_{\zeta} \delta_{\xi} ) 
    T_{\mu \nu}^{ab} = V_{\nu}^{b} 
    (-\Omega^{\lambda}  V_{\mu ; \lambda}^{a} -V_{\lambda}^{a} \tensor{\Omega}{^{\lambda}_{; \mu} }  
+          2 \tensor{S}{_{\mu \lambda}^{\beta}} V_{\beta}^{a} \Omega^{\lambda})
    + V_{\mu}^{a}  
(-\Omega^{\lambda}  V_{\nu ; \lambda}^{b} -V_{\lambda}^{b} \tensor{\Omega}{^{\lambda}_{; \nu} }  
+          2 \tensor{S}{_{\nu \lambda}^{\beta}} V_{\beta}^{b} \Omega^{\lambda}).
\end{equation}
Since $ T_{\mu \nu ; \lambda}^{ab} = V_{\mu ; \lambda}^{a} V_{\nu}^{b} + V_{\mu}^{a} V_{\nu ; \lambda}^{b}$, we have that 
\begin{equation}\label{eq:16G}
     ( \delta_{\xi} \delta_{\zeta} - \delta_{\zeta} \delta_{\xi} ) 
    T_{\mu \nu}^{ab} = 
    -\Omega^{\lambda}  T_{\mu \nu ; \lambda}^{ab} 
    - T_{\lambda \nu}^{ab} \tensor{\Omega}{^{\lambda}_{; \mu} }  
    - T_{\mu \lambda }^{a}  \tensor{\Omega}{^{\lambda}_{; \nu} }  
    +          2 \tensor{S}{_{\mu \lambda}^{\beta}} T_{\beta \nu}^{ab}  \Omega^{\lambda}
    +          2 \tensor{S}{_{\nu \lambda}^{\beta}} T_{\mu \beta}^{ab}  \Omega^{\lambda}
    ,
\end{equation}
which shows that the commutator of the variation of the tensor $ T_{\mu \nu}^{ab} $ is itself a gauge transformation of the same form as in Eq.~\eqref{eq:covarianttensor1}. 

Therefore, we can state that 
\begin{equation}\label{eq:closure}
    [ \delta_{\xi} , \delta_{\zeta} ] = \delta_{\Omega},
\end{equation}
where $ \delta $ represent a diffeomorphism and $ \Omega^{\lambda} = \delta_{\xi} \zeta^{\lambda} $. Therefore, the gauge algebra is closed for diffeomorphisms, even in the space-time with torsion. 

\section{The structure constants}\label{section:SCofGA}

In order to obtain the structure constants of the gauge algebra described by Eqs.~\eqref{eq:44}, \eqref{eq:46} and \eqref{eq:49}, let us compute the commutator of two gauge transformations using Eq.~\eqref{eq:15a}. 

First, we will consider the diffeomorphisms. We have that 
\begin{equation}\label{eq:B1}
    \delta_{\sigma} \delta_{\theta} q_{i} = \delta_{\sigma} (r_{ij} (e) \theta_{j}) = 
    \frac{\partial r_{ij} (e)}{\partial e_{k}} r_{kl} (e)\sigma_{l} \theta_{j} + r_{ij} (e) \delta_{\sigma} \theta_{j} , 
\end{equation}
where $ \delta_{\sigma} \theta_{\mu} =\mathfrak{r}_{\mu \nu} ( \theta ) \sigma^{\nu} = (- \theta_{\mu , \nu} - \theta_{\nu} \partial_{\mu}) \sigma^{\nu}  $. 
The commutator reads 
\begin{equation}\label{eq:B2}
    ( \delta_{\sigma} \delta_{\theta} - \delta_{\theta} \delta_{\sigma} ) q_{i} =
    \left(\frac{\partial r_{ij} (e)}{\partial e_{k}} r_{kl}(e)  
-\frac{\partial r_{il} (e)}{\partial e_{k}} r_{kj}(e) \right)\sigma_{l} \theta_{j} 
+ r_{ij} (e) \mathfrak{r}_{jk} ( \theta ) \sigma_{k}
 - r_{ij} (e) \mathfrak{r}_{jk} ( \sigma ) \theta_{k}.
\end{equation}
From Eq.~\eqref{eq:44}, we have that the commutator is also a diffeomorphism: $ \delta_{\Omega} q_{i} = r_{ij} (e) \Omega_{j} $. Remembering that $ \Omega_{i} = \delta_{\sigma} \theta_{i} =  \mathfrak{r}_{ij} ( \theta ) \sigma_{j} $, we obtain 
\begin{equation}\label{eq:B3}
    \left(\frac{\partial r_{ij} (e)}{\partial e_{k}} r_{kl}(e)  
-\frac{\partial r_{il} (e)}{\partial e_{k}} r_{kj}(e) \right)\sigma_{l} \theta_{j} 
=  r_{ij} (e) \mathfrak{r}_{jk} ( \sigma ) \theta_{k}.
\end{equation}
Using above relation we identity the structure constants of Eq.~\eqref{eq:27a} with 
\begin{subequations}
\label{eq:B4}
\begin{align}
    A_{km}^{l} ={}& \frac{1}{2} \frac{\partial  \tensor{\mathfrak{r} }{^{l}_{k}}( \sigma )}{\partial \sigma_{m} } \\ \intertext{and}
    B_{km}^{L} ={}& 0.
\end{align}
\end{subequations}
By Eq.~\eqref{e8b}, we find that 
\begin{equation}\label{eq:B5}
    A_{\mu \nu}^{\lambda} = \frac{1}{2} \frac{\partial}{\partial \sigma^{\nu}} \left ( - \partial_{\mu} \sigma^{\lambda} + \delta_{\mu}^{\lambda} \sigma^{\alpha} \partial_{\alpha}\right ) = \frac{1}{2} ( - \delta_{\nu}^{\lambda} \partial_{\mu} + \overleftarrow{\partial}_{\nu} \delta_{\mu}^{\lambda} ).
\end{equation}

Now, let us consider local Lorentz transformations. In this case, we obtain that 
\begin{equation}\label{eq:B6}
    \left(\delta_{\Sigma} \delta_{\Theta} - \delta_{\Theta} \delta_{\Sigma} \right)q_{i} 
    = \left(
    \frac{\partial r_{iJ} (e) }{\partial e_{k}} r_{kM} (e)- 
    \frac{\partial r_{iM} (e)}{\partial e_{k}} r_{kJ} (e)\right) 
    \Sigma_{M} \Theta_{J}= r_{iJ} (e)L^{J} ( \Theta , \Sigma ),
\end{equation}
where we have used the result in Eq.~\eqref{eq:46} which implies that $ L^{lm} ( \Theta , \Sigma ) \equiv \tensor{\Theta}{^{l}_{k}} \Sigma^{km}- \tensor{\Sigma}{^{l}_{k}} \Theta^{km}$ . Note that, Eq.~\eqref{eq:B6} is similar to Eq.~\eqref{eq:B2}. However, in Eq.~\eqref{eq:B2} we have additional terms due to fact that gauge parameters transforms under diffeomorphisms. Comparing Eq.~\eqref{eq:B6} and Eq.~\eqref{eq:27b}, we find that 
\begin{subequations}\label{eq:B7}
    \begin{align}
        A_{KM}^{l} ={}& 0 \\ \intertext{and}
    B_{ab cd}^{lm} ={}& \frac{1}{2} \frac{\partial^{2} }{\partial \Theta^{ab} \partial \Sigma^{cd}} \left ( \tensor{\Theta}{^{l}_{k}} \Sigma^{km}- \tensor{\Sigma}{^{l}_{k}} \Theta^{km}\right ) \\ \nonumber
        ={}& \frac{1}{2}  \left(\tensor{I}{^{l}_{kab}} \tensor{I}{^{km}_{cd}} -\tensor{I}{^{l}_{kcd}} \tensor{I}{^{km}_{ab}}   \right),
\end{align} 
\end{subequations}
where $ \tensor{I}{_{ab}^{cd}} = ( \delta_{a}^{c} \delta_{b}^{d} - \delta_{a}^{d} \delta_{b}^{c} )/2$ is the antisymmetric identity.

Finally, let us consider mixed transformations. In this case, we have 
\begin{subequations}\label{eq:B8}
\begin{align}\label{eq:B8a}
    \delta_{\Theta} \delta_{\theta} q_{i} ={}& \frac{\partial r_{ij} (e) }{\partial e_{k}} r_{kM} \Theta_{M} \theta_{j}  
    \\
    \intertext{and}
    \delta_{\theta} \delta_{\Theta} q_{i} ={}& \frac{\partial r_{iM} (e) }{\partial e_{k}} r_{kj} (e)\Theta_{M} \theta_{j}  + r_{iM} (e) \mathfrak{r}_{Mj} ( \Theta )\theta_{j}.
\end{align}
\end{subequations}
By Eq.~\eqref{eq:49}, we find that 
\begin{equation}\label{eq:B9}
    \left(
    \frac{\partial r_{ij} (e) }{\partial e_{k}} r_{kM}(e)- 
    \frac{\partial r_{iM} (e) }{\partial e_{k}} r_{kj}(e)
    \right)
    \Theta_{M} \theta_{j} =  r_{iM} (e) \mathfrak{r}_{Mj} ( \Theta ) \theta_{j} . 
\end{equation}
Comparing Eq.~\eqref{eq:B9} with Eq.~\eqref{eq:27c}, we get that 
\begin{subequations}\label{eq:B10}
    \begin{align}
        A_{kM}^{l} ={}& 0 \\ \intertext{and}
        B_{\mu ab}^{lm} ={}&\frac{\partial \tensor{\mathfrak{r}}{^{lm}_{\mu }}( \Theta )}{\partial \Theta^{ab} } = -\tensor{I}{^{lm}_{ab}} \partial_{\mu}, 
    \end{align}
\end{subequations}
where $ \tensor{\mathfrak{r}}{^{lm}_{\mu}} ( \Theta ) \equiv  -\partial_{\mu} \Theta^{lm} $. 

Alternatively, by replacing in Eq.~\eqref{eq:25} $r$, $ \phi $, $A$, $B$ by $R$, $ \Phi $, $A'$, $B'$; one would obtain the following relation which solves the Eq.~\eqref{eq:26b}
\begin{subequations}\label{eq:aB17b}
    \begin{align} \label{eq:aB17ab}
        \frac{1}{2} \left ( \frac{\partial R_{Jk}}{\partial \Phi_{L}} R_{Lm} - \frac{\partial R_{Jm}}{\partial \Phi_{L}} R_{Lk}\right ) ={}& R_{Jl} A_{km}^{\prime l} + R_{JL} B_{km}^{\prime L},\\ \label{eq:aB17bb}
        \frac{1}{2}\left ( \frac{\partial R_{JK}}{\partial \Phi_{L}} R_{LM} - \frac{\partial R_{JM}}{\partial \Phi_{L}} R_{LK}\right ) ={}& R_{Jl} A_{KM}^{\prime l} + R_{JL} B_{KM}^{\prime L}, \\ \label{eq:aB17cb}
        \frac{\partial R_{Jk}}{\partial \Phi_{L}} R_{LM} - \frac{\partial R_{JM}}{\partial \Phi_{L}} R_{Lk} ={}& R_{Jl} A_{kM}^{\prime l} + R_{JL} B_{kM}^{\prime L}.
    \end{align}
\end{subequations}
We now remark that, with the replacements $q$, $e$ $ \rightarrow$ $Q$, $ \omega $, the Eqs.~\eqref{eq:44}, \eqref{eq:46} and \eqref{eq:49} become similar to the Eqs.~\eqref{eq:51a}, \eqref{eq:51b} and \eqref{eq:52}. Thus, by replacing in Eqs.~\eqref{eq:B3}, \eqref{eq:B6} and \eqref{eq:B9}, $r$, $q_i$ and $e_k$ with $R$, $Q_I$ and $ \omega_{K} $, and proceeding as in the previous case, one finds that $A'=A$ and $B'=B$.

The uniqueness of the matrices $A$ and $B$ imply certain structural relations, which may be more simply understood by comparing the 
Eqs.~\eqref{eq:15} and \eqref{eq:30}. This allows to identity the matrix elements of $r$ and $R$ as: 
\begin{align} \label{eq:c3}
    r_{\mu \sigma}^{a} (e) ={}& - e^{a}_{\mu , \sigma} - e_{\sigma}^{a} \partial_{\mu} ; \quad r_{\mu a}^{lm} (e) = I_{ac}^{lm} e_{\mu}^c, \\ \label{eq:c4}
    R_{\mu ab \sigma } ( \omega ) ={}&- \omega_{\mu ab , \sigma} - \omega_{\sigma ab} \partial_{\mu} ; \quad R_{\mu ab }^{lm} (\omega )= 
    \frac{1}{2} {I}{_{ab}^{lm}} \partial_{\mu} +   {I}{_{ac}^{lm}} \omega_{\mu\; b}^{\; c\;}
          - (a \leftrightarrow b),
\end{align}
where $ I_{ab}^{cd} = ( \delta_{a}^{c} \delta_{b}^{d} - \delta_{a}^{d} \delta_{b}^{c}  )/2$ is the identity matrix. 

We note here that the functional derivatives of $r (e)$ with respect to $e$, and of $R( \omega )$ with respect to $ \omega $, are closely related. Such derivative matrix elements, which occur in the Eq.~\eqref{eq:27} and \eqref{eq:aB17b}, uniquely determine the matrices $A$ and $B$.

\subsection{BRST transformations of the ghosts fields}\label{section:BRST}

Employing the results found in this appendix, we find that 
\begin{subequations}\label{eq:BRSTap}
    \begin{align} \label{eq:BRST11}
        \delta c^{\mu} ={}&- c^{\alpha} A_{\alpha \beta}^{\mu}  c^{\beta} \eta  = c^{\alpha} \tensor{c}{^{\mu}_{, \alpha }} \eta ,
        \\ \intertext{and} \label{eq:BRST22}
        \delta C^{ab} ={}& - C^{lm} B^{ab}_{lm pq}  C^{pq} \eta - c^{\mu} B_{\mu  lm}^{ab} C^{lm} \eta =\left( -\tensor{C}{^{a}_{k}} C^{kb}   + c^{\mu} \tensor{C}{^{ab}_{, \mu}} \right)\eta  .
    \end{align}
\end{subequations}
We can rewrite Eq.~\eqref{eq:BRST11} in a covariant way by using that 
\begin{equation}\label{eq:B12}
    c^{\alpha} \tensor{c}{^{\mu}_{, \alpha }} = c^{\alpha} (\tensor{c}{^{\mu}_{; \alpha}} - \tensor{\Gamma}{_{\alpha \beta}^{\mu}} c^{\beta} ) = c^{\alpha} \tensor{c}{^{\mu}_{; \alpha}} - \tensor{S}{_{\alpha \beta}^{\mu}} c^{\alpha} c^{\beta},
\end{equation}
where we used that $c^{\alpha} $ is an anticommuting field. 

We can show that the transformations \eqref{eq:BRSTap} are nilpotent provided that \cite{Nishijima:1978wq} 
\begin{equation}\label{eq:B13}
    [\delta, \partial_{\mu}] = 0. 
\end{equation}
Consider the transformation of Eq.~\eqref{eq:BRST11}: 
\begin{equation}\label{eq:B14a}
    \delta ( c^{\alpha} \tensor{c}{^{\mu}_{, \alpha}})  = -(c^{\beta} \tensor{c}{^{\alpha}_{, \beta} }  )\tensor{c}{^{\mu}_{, \alpha}} \eta 
+c^{\beta} ( c^{\alpha} \tensor{c}{^{\mu}_{,\alpha}} )_{, \beta} \eta ,
\end{equation}
where we used Eq.~\eqref{eq:B13}. Now, using the Grassmann character of $ c^{\mu} $, we find that 
\begin{equation}\label{eq:B15a}
     \delta ( c^{\alpha} \tensor{c}{^{\mu}_{, \alpha}})  =  c^{\beta}  c^{\alpha} \tensor{c}{^{\mu}_{,\alpha \beta }} \eta  
     =0
\end{equation}
which shows that the transformation \eqref{eq:BRST1} is nilpotent.
The nilpotency of the transformation in Eq.~\eqref{eq:BRST22} can also be verified. Instead, we will consider the transformation: 
\begin{equation}\label{eq:B16a}
    \begin{split}
        \delta ( -\alpha \tensor{C}{^{a}_{k}} C^{kb} + \beta c^{\mu} \tensor{C}{^{ab}_{, \mu}}) ={}&  \alpha \left ( -\alpha \tensor{C}{^{a}_{l}} \tensor{C}{^{l}_{k}} + \beta c^{\mu} \tensor{C}{^{a}_{k , \mu}}\right ) C^{kb} \eta - \alpha \tensor{C}{^{a}_{k}} \left ( -\alpha \tensor{C}{^{k}_{l}} C^{lb} + \beta c^{\mu} \tensor{C}{^{kb}_{, \mu}} \right )  \eta \\ & \quad -\beta c^{\nu} \tensor{c}{^{\mu}_{,\nu}} \tensor{C}{^{ab}_{, \mu}}   \eta + 
    \beta c^{\nu} \left ( -\alpha \tensor{C}{^{a}_{k}} C^{kb} + \beta c^{\mu} \tensor{C}{^{ab}_{,\mu}}\right )_{,\nu} \eta 
    \\
        ={}&
        \beta ( \beta -1) c^{\nu} \tensor{c}{^{\mu}_{, \nu }} \tensor{C}{^{ab}_{, \mu}} \eta .
\end{split}
\end{equation}
Thus, we find that the transformation Eq.~\eqref{eq:BRST22} is nilpotent ($ \alpha =\beta = 1 $). It is also possible to obtain other nilpotent transformations, e.g., by setting $ \beta = 0 $ and $ \alpha =1$, we have $ \delta C^{ab} = \delta_{\text{LT}} C^{ab} =-\tensor{C}{^{a}_{k}} C^{kb} \eta$. We also can set $ \alpha =0 $ and $ \beta = 1$ that leads to $ \delta C^{ab} = \delta_{\text{Diff}} C^{ab} = c^{\mu} \tensor{C}{^{ab}_{, \mu}} \eta$.  However, in such cases $ \delta^{2} e_{\mu}^{a} $ and $ \delta^{2} \omega_{\mu ab} $ no longer vanish.

\subsection{Closure of the gauge algebra for ghost fields}\label{section:CGAGF}

We point out that similarly to Eq.~\eqref{eq:closure}, we also have that 
\begin{align}\label{eq:B14}
    [ \delta_{\lambda} , \delta_{\rho} ] ={}& \delta_{L},\\
    [ \delta_{\lambda} , \delta_{\xi} ] ={}& 0;
\end{align}
where $ \delta_{\lambda} $ and $ \delta_{\rho} $ are local Lorentz transformations and $ \delta_{\xi} $ is a diffeomorphism. The parameter $ \tensor{L}{^{a}_{b}} = \delta_{\lambda} \tensor{\rho}{^{a}_{b}}$ is defined in Eq.~\eqref{eq:46}. 
It may be verified that these relations holds for the ghosts fields $ c^{\mu} $ and $ C^{ab} $ or any other field. This shows that the gauge algebra of the EC theory is closed and that it separates into two decoupled parts.

For example, let us compute $ [ \delta_{\lambda} , \delta_{\xi} ] C^{ab} $. We have that 
\begin{equation}\label{eq:B15}
    \delta_{\xi} C^{ab} = - \xi^{\sigma} \tensor{C}{^{ab}_{, \sigma}} 
\end{equation}
and that 
\begin{equation}\label{eq:B16}
    \delta_{\lambda } C^{ab} = \tensor{\lambda}{^{a}_{d} } C^{d b} + \tensor{\lambda}{^{b}_{d}} C^{ad}. 
\end{equation}
Hence, we find that
\begin{equation}\label{eq:B17}
    [ \delta_{\lambda } , \delta_{\xi} ] C^{ab} = - \xi^{\sigma}  (\tensor{\lambda}{^{a}_{d} } C^{d b} + \tensor{\lambda}{^{b}_{d}} C^{ad})_{, \sigma} + \xi^{\sigma} \tensor{\lambda}{^{a}_{d , \sigma}} C^{db} + \xi^{\sigma} \tensor{\lambda}{^{a}_{d}} \tensor{C}{^{db}_{, \sigma}} + \xi^{\sigma} \tensor{\lambda}{^{b}_{d, \sigma}} C^{ad} + \xi^{\sigma} \tensor{\lambda}{^{b}_{d}} \tensor{C}{^{ad}_{,\sigma}} =0.
\end{equation}
This relation is similar to that given in Eqs.~\eqref{eq:49} and \eqref{eq:52}.

\end{document}